\documentclass[reprint,aps,prl,amsmath,amssymb,amsfonts,dvips]{revtex4-1}
\usepackage[dvips]{graphicx,color}
\usepackage{braket,bm}
\usepackage{dcolumn}
\begin{document}

\title{Monte Carlo study of the critical properties of noncollinear Heisenberg magnets: \\ $O(3)\times O(2)$ universality class}

\author{Yoshihiro Nagano, Kazuki Uematsu and Hikaru Kawamura}
\email[]{kawamura@ess.sci.osaka-u.ac.jp}
\affiliation{Department of Earth and Space Science, Graduate School of Science, Osaka University, Toyonaka, Osaka 560-0043, Japan}

\date{\today}

\begin{abstract}
The critical properties of the antiferromagnetic Heisenberg model on the three-dimensional stacked-triangular lattice are studied by means of a large-scale Monte Carlo simulation in order to get insight into the controversial issue of the criticality of the noncollinear magnets with the $O(3)\times O(2)$ symmetry. The maximum size studied is $384^3$, considerably larger than the sizes studied by the previous numerical works on the model. Availability of such large-size data enables us to examine the detailed critical properties including the effect of corrections to the leading scaling. Strong numerical evidence of the continuous nature of the transition is obtained. Our data indicates the existence of significant corrections to the leading scaling. Careful analysis by taking account of the possible corrections yield critical exponents estimates, $\alpha=0.44(3)$, $\beta=0.26(2)$, $\gamma=1.03(5)$, $\nu=0.52(1)$, $\eta=0.02(5)$, and the chirality exponents $\beta_\kappa=0.40(3)$ and $\gamma_\kappa=0.77(6)$, supporting the existence of the $O(3)$ chiral (or $O(3)\times O(2)$) universality class governed by a new `chiral' fixed point. We also obtain an indication that the underlying fixed point is of the focus-type, characterized by the complex-valued correction-to-scaling exponent, $\omega=0.1^{+0.4}_{-0.05} + i\ 0.7^{+0.1}_{-0.4}$. The focus-like nature of the chiral fixed point accompanied by the spiral-like renormalization-group (RG) flow is likely to be the origin of the apparently complicated critical behavior. The results are compared and discussed in conjunction with the results of other numerical simulations, several distinct types of RG calculations including the higher-order perturbative massive and massless RG calculations and the nonperturbative functional RG calculation, and  the conformal-bootstrap program.
\end{abstract}

%\pacs{}

\maketitle

\section{I. Introduction}

 The concept of universality has been a cornerstone of modern theory of phase transition and critical phenomena. According to the universality hypothesis, critical properties associated with continuous phase transitions possess universal features independent of microscopic details of each system, and can be classified into a small number of universality classes. Each universality is specified by the symmetry of the order parameter, the spatial dimensionality and the range of interaction. 

 Magnetic systems have offered a framework for the study of the critical phenomena and the universality class for years. In bulk magnets, the universality class is usually labeled by the number of spin components $n$, {\it i.e.\/}, $n=1$ (Ising), $n=2$ ($XY$) and $n=3$ (Heisenberg) depending on whether the interaction is easy-axis type, easy-plane-type or isotropic, respectively. % Sometimes, modification to this classification occurs, {\it e.g.\/}, the cubic anisotropy giving rise to a different universality class, The modification is relatively minor, in the sense that the exponents deviate not much from those of the isotropic Heisenberg class.

 In the middle eighties, one of the present authors (H.K.) suggested on the basis of a symmetry argument, Monte Carlo (MC) simulations and renormalization-group (RG) analysis that certain {\it frustrated\/} magnets with the {\it noncollinear\/} spin order might exhibit a phase transition belonging to a new universality class, $O(n)$ `chiral' universality class, different from the well-known $O(n)$ universality class \cite{KawamuraMCH,KawamuraMCXY,KawamuraRG,KawamuraRG2,KawamuraMCHXY,Kawamura-review}. We begin with a summary of these earlier works. Concerning a symmetry, the order-parameter space $V$ isomorphic to the set of ordered state of the frustrated noncollinear $n$-component magnets is $O(n)/O(n-2)$ \cite{KawamuraMCH,KawamuraMCXY,Kawamura-review}, instead of $O(n)/O(n-1)=S_{n-1}$ ($S_n$ the $n$-dimensional sphere) of the collinear order in standard unfrustrated $n$-component magnets. The associated Landau-Ginzburg-Wilson (LGW) Hamiltonian can be written in terms of two $n$-component vector fields, in contrast to a single $n$-component field in the standard $n$-component $\phi^4$ model, with the associated symmetry $O(n)\times O(2)$, in contrast to $O(n)$ of the standard  $\phi^4$ model \cite{KawamuraRG,Kawamura-review}. Then, $O(n)$ chiral universality class is sometimes called $O(n)\times O(2)$ universality class. This symmetry was further extended to $O(n)\times O(m)$ \cite{KawamuraRG2,Kawamura-review}. Renormalization-group (RG) analysis based the LGW Hamiltonian including both the $\epsilon=4-d$ and $1/n$ expansions were performed, to yield a new `chiral' fixed point (FP) for larger $n$ \cite{KawamuraRG,KawamuraRG2}. More precisely, the second-order $\epsilon$-expansion yielded the stability region of the chiral FP to be $n\geq n_c(\epsilon)=21.8-23.4\epsilon + O(\epsilon^2)$. Whether the physically relevant case of $d=3$ and $n=2$, 3 is included in this region or not has been not so clear, however. Concerning the Monte Calro (MC) study of microscopic spin models, the MC simulations of Refs.\cite{KawamuraMCH,KawamuraMCXY,KawamuraMCHXY} studied the classical vector ($n=2$ or 3) antiferromagnet on the three-dimensional (3D) stacked-triangular lattice, observing a continuous transition. In the Heisenberg ($n=3$) case, the exponents were estimated to be $\alpha=0.24(8)$, $\beta=0.30(2)$, $\gamma=1.17(7)$, $\nu=0.59(2)$ and $\eta=0.02(18)$ \cite{KawamuraMCHXY}, where $\alpha,\ \beta,\ \gamma,\ \nu,\ \eta$ are the specific-heat, the order-parameter, the ordering-susceptibility, the correlation-length and the critical-point-decay exponents, respectively. The chiral exponents were also estimated to be $\beta_\kappa=0.55(4)$ and $\gamma_\kappa=0.72(8)$ \cite{KawamuraMCHXY} where $\beta_\kappa$ and $\gamma_\kappa$ are the chirality and the chiral-susceptibility exponents, respectively. A series of these earlier theoretical works by one of the present authors was reviewed in Ref.\cite{Kawamura-review}

 Since then, a lot of both theoretical and experimental activities have been made on the noncollinear criticality of frustrated magnets. Some support the existence of a new universality class, while some others suggest the absence of a new universality class claiming the noncollinear transition being first order. 

 Concerning MC and related numerical simulations, earlier MC simulations on the stacked-triangular AF Heisenberg model, the same model as studied in Refs.\cite{KawamuraMCH,KawamuraMCXY,KawamuraMCHXY} yielded a continuous transition characterized the exponents more or less similar to the ones reported in Ref.\cite{KawamuraMCHXY}, including the works by Bhattacharya {\it et al\/} \cite{Bhattacharya}, Mailhot {\it et al\/} \cite{Mailhot}, and Loison {\it et al\/} \cite{Loison}, though the lattice sizes studied were rather small $L\leq 32-48$. By contrast, Itakura performed the MCRG study of the LGW model on the lattice, and concluded that the transition was of first order for both cases of $n=2$ and $n=3$ \cite{Itakura}. Ngo and Diep applied the Wang-Landau method to the AF Heisenberg model on the stacked-triangular lattice, exactly the same model as studied in Refs.\cite{KawamuraMCH,KawamuraMCXY,KawamuraMCHXY}, and concluded that the transition was actually first order based on the observation of the double peaks in the energy distribution at $T_c$ \cite{Diep}. They argued that the system size studied in the previous simulations were too small to unambiguously identify the first-order transition. Hence, the numerical situation on the phase transition of the AF Heisenberg model on the stacked-triangular lattice has remained unclear.

 The situation of the RG analysis has also remained unclear. The two-loop analysis of Ref.\cite{KawamuraRG} was extended to three-loop order by Antonenko and Sokolov \cite{Antonenko1}. By applying the Pad\'e-Borel analysis, these authors concluded that the chiral FP did not exist for the physically relevant cases of $n=2$ and 3 in 3D, and the transition was first-order. The three-loop $\epsilon$-expansion calculation found the chiral fixed point stabilized at $n\geq n_c=21.8-23.4\epsilon+7.1\epsilon^2+O(\epsilon^3)$ \cite{Antonenko2}. Subsequently, higher-order perturbative RG calculations in fixed $d=3$ dimensions combined with the resummation technique were performed by Pelissetto, Vicari, Calabrese and collaborators based on the two distinct RG schemes, {\it i.e.\/}, six-loop calculation with the massive zero momentum (MZM) scheme \cite{Pelissetto1,Pelissetto2,Calabrese1} and the five-loop calculation with the massless minimal subtraction ($\overline{{\rm MS}}$) scheme \cite{Calabrese2}. In contrast to the lower-order three-loop calculation \cite{Antonenko1,Antonenko2}, both schemes lead to the stable chiral FP associated with a continuous transition at $d=3$ both for $n=2$ and 3. Interesting observation here is that the chiral FP was of the peculiar ``focus-type'' FP with a {\it complex-valued\/} correction-to-scaling exponent, where the RG flow exhibits a spiral-like flow into the chiral FP \cite{Calabrese1,Calabrese2}. The estimated exponents differ somewhat between the two RG schemes, {\it i.e.\/}, $\alpha=0.35(9)$, $\beta=0.30(2)$, $\gamma=1.06(5)$, $\nu=0.55(3)$, $\eta=0.073(94)$ \cite{Pelissetto1}, $\beta_\kappa=0.38(10)$ and $\gamma_\kappa=0.89(10)$ \cite{Pelissetto2} in the massive MZM scheme, and $\alpha=0.11(15)$, $\beta=0.34(3)$, $\gamma=1.20(8)$, $\nu=0.63(5)$, $\eta=0.08(3)$, $\beta_\kappa=0.54(17)$ and $\gamma_\kappa=0.81(23)$ in the massless $\overline{{\rm MS}}$ scheme \cite{Calabrese2}.

 In sharp contrast, on the basis of a series of nonperturbative functional RG calculations, Delamotte, Tisser, Mouhana and collaborators claimed that the $O(n)\times O(2)$ model did not possess any new fixed point and the transition of noncollinear magnets should be first order \cite{Tisser1,Tisser2,Delamotte1,Delamotte2,Delamotte3}. While this RG scheme is nonperturbative, it contains some approximation/truncation whose validity is not totally clear. Hence, the RG situation has remained quite controversial. Since both the higher-order perturbative approach and the non-perturbative approach give consistent results in the standard cases of the $O(n)$ Heisenberg model, the cause of the observed sharp discrepancy between the two RG methods remains to be understood.

 More recently, still another theoretical approach, the conformal bootstrap program, was applied to this problem by Nakayama and Ohtsuki \cite{Nakayama}. The method imposes the ``exact'' bound to the scaling dimensions of operators. A kink-like singular behavior is sometimes realized in the bounds, which is employed to give quite accurate estimates of critical exponents.
%The method does not assume any Hamiltonian, neither the Heisenberg model nor the LGW Hamiltonian, but just the symmetry $O(n)\times O(2)$ and some fundamental assumptions of conformal covariance and reflection positivity. 
In the $n=3$ case, the program leads to a continuous transition in $d=3$, with the exponents estimates $\alpha=0.10(6)$, $\beta=0.34(1)$, $\gamma=1.22(4)$, $\nu=0.63(2)$, $\eta=0.078(6)$, $\beta_\kappa=0.56(7)$ and $\gamma_\kappa=0.77(10)$ \cite{Nakayama}. The obtained values turned out to be rather close to the estimates of the high-order perturbative massless RG \cite{Calabrese2}. Since the conformal bootstrap approach is completely independent of and different from the RG approaches, this result seems to strengthen the existence of the $O(3) \times O(2)$ universality class. Meanwhile, the conformal-bootstrap program assumes the absence of the focus point suggested from the higher-order perturbative RG \cite{Calabrese1,Calabrese2}, and the situation still remains not totally clear.

 Under such circumstances, in order to get further insights into the issue, we wish to perform in the present paper a large-scale MC simulation on the AF Heisenberg model on the 3D stacked-triangular lattice. The model is the same one as studied previously in Refs.\cite{KawamuraMCH,KawamuraMCHXY,Bhattacharya,Mailhot,Loison,Diep}, but here we go to lattices considerably larger than those studied before, {\it i.e.\/}, up to $N=L^3=384^3$. By so doing, we wish to perform more precise analysis of the critical properties than before. Indeed, we give a rather precise estimate of the transition temperature $T_c=0.957270\pm 0.000004$, and find strong numerical evidence that the transition is continuous. We also find significant corrections to the leading scaling. By carefully examining the correction-to-scaling effects, we get the estimates of critical exponents, $\alpha=0.44(3)$, $\beta=0.26(2)$, $\gamma=1.03(5)$, $\nu=0.52(1)$, $\eta=0.02(5)$, and the chirality exponents $\beta_\kappa=0.40(3)$ and $\gamma_\kappa=0.77(6)$, Quite interestingly, we find some indication of the focus-type FP, {\it i.e.\/}, the complex-valued correction-to-scaling exponent $\omega=0.1^{+0.4}_{-0.05} + i\ 0.7^{+0.1}_{-0.4}$

 The rest of the present paper is organized as follows. In Section II, we introduce the model and the numerical method employed. Computed physical quantities in our MC simulations are defined and some of their properties are explained in Section III. In Section IV, we show the MC data of representative physical quantities in the transition region. Section V is devoted to the precise determination of the transition temperature $T_c$, and the order of the transition is examined in Section VI. Section VII consists of thee subsections, and is devoted to the estimates of various critical exponents. The analysis without the correction term is first given in Section VII-1. The analysis invoking one and two real correction-to-scaling exponents are made in Section VII-2, while that invoking a complex correction-to-scaling exponent is made in Section VII-3. Finally, Section VIII is devoted to summary and discussion. In Appendix, we derive the general expression of the exponent describing the size dependence of the energy Binder ratio at a continuous transition.

\section{II. The model and the method}

Our model is the classical Heisenberg model on the 3D stacked-triangular or simple-hexagonal lattice with the antiferromagnetic (AF) nearest-neighbor (NN) interaction, whose Hamiltonian is given by
\begin{equation}
{\cal H} = J \sum_{\langle ij \rangle} {\bf S}_i\cdot {\bf S}_j,
\end{equation}
where ${\bf S_i}=(S_i^x, S_i^y, S_i^z)$ ($|{\bf S_i}|=1$) is the three-component unit vector at the $i$-th site, $J>0$ is the NN AF coupling, and the summation  $\langle ij \rangle$ is taken over all NN pairs on the stacked-triangular lattice including both intra- and inter-plane bonds.  Following the earlier numerical works \cite{KawamuraMCH,KawamuraMCHXY,Bhattacharya,Mailhot,Loison}, we assume for simplicity that the intra- and inter-plane interactions are of the same magnitude $J$. In the present paper, we take the energy (the temperature) unit of $J=1$. The lattice consists of $N=L^3$ sites, periodic boundary conditions applied in all directions.

 Thermodynamic properties of the model are investigated by means of MC simulations based on the standard heat-bath method combined with the over-relaxation method. One MC step per spin (MCS) consists of one heat-bath sweep followed by ten successive over-relaxation sweeps.

 The lattice sizes studied are $L=12$, 18, 24, 30, 36, 48, 54, 60, 72, 96, 108, 120, 144, 192, 240, 288 and 384. The largest lattice size studied $L=384$ is significantly greater than those previously studied on the same model, {\it i.e.\/}, $L\leq 60$ by MC \cite{KawamuraMCH,KawamuraMCHXY,Bhattacharya,Mailhot,Loison} and $L\leq 150$ by the Wang-Landau method \cite{Diep}.

 Equilibration is checked by monitoring the MC-time dependence of physical quantities: See also Section III below. Typically, after discarding initial $10^6$ MCS for equilibration, subsequent $2.5\times 10^6$ MCS are used to compute thermal averages of physical quantities. At each temperature and lattice size, twelve independent runs are made with using different spin initial condition and different random-number sequences. Error bars are estimated from the distribution of the data over these twelve independent runs.

 Since our interest in the present paper concerns with the critical properties, we focus on the thermodynamic properties in the temperature range close to the transition temperature $T_c$. Long MC runs are made at a specific temperature or at several specific temperatures close to $T_c$, and thermodynamic properties at nearby temperatures are obtained by use of the histogram technique \cite{Ferrenberg}. We restrict the range of the temperature shift from the original temperature at which the data are taken to the temperature where the shifted energy distribution has a considerable overlap with the original distribution. More precisely, let $e$ the energy per spin and $P(e)$ the energy distribution. When the original energy distribution $P(e)$ takes values greater than the half of its peak value in the energy range between $e_L$ and $e_R$, we limit the shifted temperature so that the peak position of the shifted $P(e)$ lies in the range [$e_L,e_R$]. Most of the data are taken at $T=0.95727$, our best estimate of the transition temperature $T_c$ to be determined below, whereas some data are taken at other nearby temperatures for the consistency check.

\section{III. Physical quantities}

In this section, we introduce various physical quantities we compute by MC. The internal energy per spin $\bar e$ is the thermal average of the Hamiltonian normalized by the total number of the spin, $\bar e=\langle e \rangle=\langle {\cal H}\rangle /N$, where $\langle \cdots \rangle$ denotes the thermal average. As mentioned, all the energy and the temperature have been normalized by $J$. The specific heat per spin $c$, measured in units of $k_B$, is calculated from the energy fluctuation. The energy Binder ratio \cite{Binder} $g_e$ is defined by
\begin{equation}
g_e=\frac{\langle e^4\rangle}{\langle e^2\rangle^2}.
\end{equation}

The model is known to exhibit the AF long-range order (LRO) in the ordered state, taking the $120^\circ$ spin structure. We define the corresponding AF order parameter $m_{AF}$ via an appropriate spin Fourier component ${\bm S}({\bm Q})$,
\begin{equation}
m_{AF}=\langle |{\bm S}({\bm Q})| \rangle,\ \  
{\bm S}({\bm Q})=\frac{1}{N} \sum_i {\bm S}_ie^{i{\bm Q}\cdot {\bm r}_i} ,
\end{equation}
where ${\bm Q}=({\frac{4\pi}{3}},0,\pi)$ is the wavevector representing the $120^\circ$ structure (the lattice constant is taken as the length unit here), and the summation over $i$ is taken over all spins on the lattice. Its temperature derivative $\frac{{\rm d}m_{AF}}{{\rm d}T}$ can be computed from 
\begin{equation}
\frac{{\rm d}m_{AF}}{{\rm d}T}=\frac{N}{T^2}(\langle e\times m_{AF}\rangle - \langle e\rangle \langle m_{AF}\rangle).
\end{equation}
We also define the associated AF susceptibility $\chi_{AF}$  by
\begin{equation}
\chi_{AF} = \frac{N}{T} \langle |{\bm S}({\bm Q})|^2 \rangle.
\end{equation}
The spin Binder ratio $g_s$ associated with the AF order is defined by 
\begin{equation}
g_s=4-3\frac{\langle m_{AF}^4\rangle}{\langle m_{AF}^2\rangle^2},
\end{equation}
where we have used, in appropriately normalizing $g_s$, the fact that the number of independent components of the AF order parameter is six, {\it i.e.\/}, three (the number of spin components) times two (the number of independent Fourier modes, ${\bm Q}$ and $-{\bm Q}$). Its temperature derivative $\frac{{\rm d}g_s}{{\rm d}T}$ can be computed from
\begin{equation}
\frac{{\rm d}g_s}{{\rm d}T}=\frac{3N}{T^2}\left( \frac{2\langle m_{AF}^4\rangle\langle e m_{AF}^2\rangle}{\langle m_{AF}^2\rangle^3} -  \frac{\langle e m_{AF}^4\rangle + \langle e\rangle \langle m_{AF}^4\rangle}{\langle m_{AF}^2\rangle^2} \right).
\end{equation}
The finite-size spin-correlation lengths are defined both for the intraplane ($\parallel$) and interplane ($\perp$) correlations by
\begin{eqnarray}
\xi^\parallel_s = \frac{1}{2\sin (\pi/L)} \sqrt{\frac{|{\bm m}({\bm Q})|^2}{|{\bm m}({\bm Q}+\delta {\bm Q}^\parallel)|^2} - 1}, \\
\xi^\perp_s = \frac{1}{2\sin (\pi/L)} \sqrt{\frac{|{\bm m}({\bm Q})|^2}{|{\bm m}({\bm Q}+\delta {\bm Q}^\perp)|^2} - 1} ,
\end{eqnarray}
where $\delta {\bm Q}^\parallel = (\frac{2\pi}{L},0,0)$ and $\delta {\bm Q}^\perp=(0,0,\frac{2\pi}{L})$ are the possible minimum nonzero wavevectors along the intra- and inter-triangular-layer directions, respectively. Although a common criticality is expected for $\xi^\parallel_s$ and $\xi^\perp_s$, we compute both quantities below. The dimensionless quantity called the correlation-length ratio, $\xi_s^\parallel/L$ or $\xi_s^\perp/L$, plays an important role in the study of critical properties. 

 The local vector chirality ${\bm \kappa}_\triangle$ may be defined for three spins on an elementary upward triangle $\triangle$ on the triangular layer by
\begin{equation}
{\bm \kappa}_\triangle = \frac{3}{2\sqrt{2}} \sum_{\langle ij\rangle \in \triangle} {\bm S}_i\times {\bm S}_j, 
\end{equation}
where the summation is taken over three NN bonds on each upward triangle in a clockwise direction. The total chirality $\kappa$ is then defined by
\begin{eqnarray}
\kappa = \langle |{\bm \kappa}| \rangle, \\
{\bm \kappa}=\frac{1}{N} \sum_\triangle {\bm \kappa}_\triangle , 
\end{eqnarray}
where the summation is taken over all $N$ upward triangles on the lattice.  Its temperature derivative $\frac{{\rm d}\kappa}{{\rm d}T}$ can be computed from 
\begin{equation}
\frac{{\rm d}\kappa}{{\rm d}T}=\frac{N}{T^2}(\langle e \kappa \rangle - \langle e\rangle \langle \kappa\rangle).
\end{equation}
The associated chiral susceptibility is defined by
\begin{equation}
\chi_\kappa = \frac{N}{T} \langle {\bm \kappa}^2\rangle.
\end{equation}
The {\it chiral\/} Binder ratio $g_\kappa$ is defined by
\begin{equation}
g_\kappa=\frac{1}{2} \left(5-3\frac{\langle \kappa^4 \rangle}{\langle \kappa^2 \rangle^2} \right).
\end{equation}
where in normalizing $g_\kappa$ we have used the fact that the number of independent components of the vector chirality is three. Its temperature derivative $\frac{{\rm d}g_\kappa}{{\rm d}T}$ can be computed from
\begin{equation}
\frac{{\rm d}g_\kappa}{{\rm d}T}=\frac{3N}{2T^2}\left( \frac{2\langle {\bm \kappa}^4\rangle\langle e{\bm \kappa}^2\rangle}{\langle {\bm \kappa}^2\rangle^3} -  \frac{\langle e{\bm \kappa}^4\rangle + \langle e\rangle \langle {\bm \kappa}^4\rangle}{\langle {\bm \kappa}^2\rangle^2} \right).
\end{equation}%
The finite-size {\it chiral\/}-correlation lengths are defined both for the intraplane and interplane correlations by
\begin{eqnarray}
\xi^\parallel_\kappa = \frac{1}{2\sin (\pi/L)} \sqrt{\frac{|{\bm \kappa}({\bm Q})|^2}{|{\bm \kappa}({\bm Q}+\delta {\bm Q}^\parallel)|^2} - 1} \\
\xi^\perp_\kappa = \frac{1}{2\sin (\pi/L)} \sqrt{\frac{|{\bm \kappa}({\bm Q})|^2}{|{\bm \kappa}({\bm Q}+\delta {\bm Q}^\perp)|^2} - 1} .
\end{eqnarray}
where ${\bm \kappa}({\bm Q})$ is the Fourier transform of ${\bm \kappa}_\triangle$,
\begin{equation}
{\bm \kappa}({\bm Q})=\frac{1}{N} \sum_\triangle {\bm \kappa}_\triangle e^{i{\bm Q}\cdot {\bm r}_\triangle},
\end{equation}
${\bm r}_\triangle$ being the position vector of the elementary triangle $\triangle$.

\begin{figure}[t]
  \begin{center}
    \includegraphics[width=40mm]{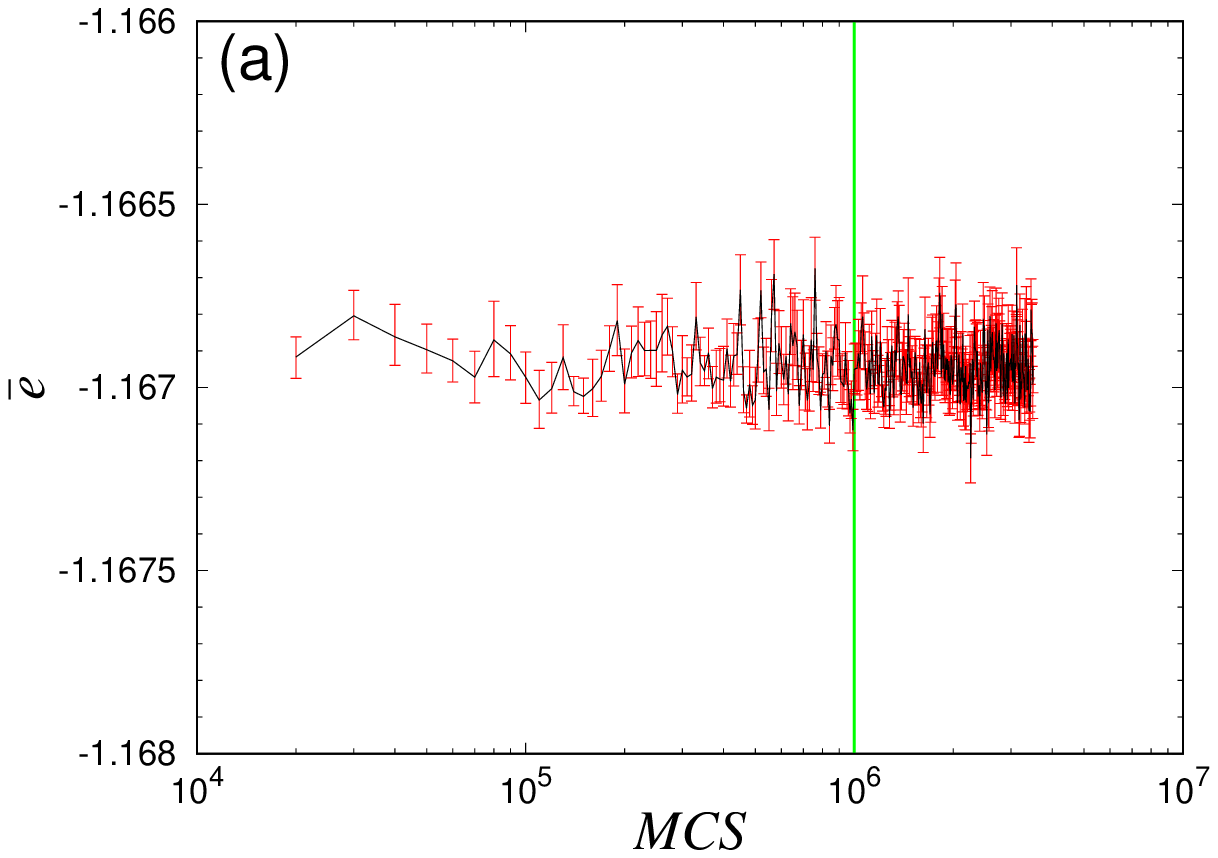}
    \includegraphics[width=40mm]{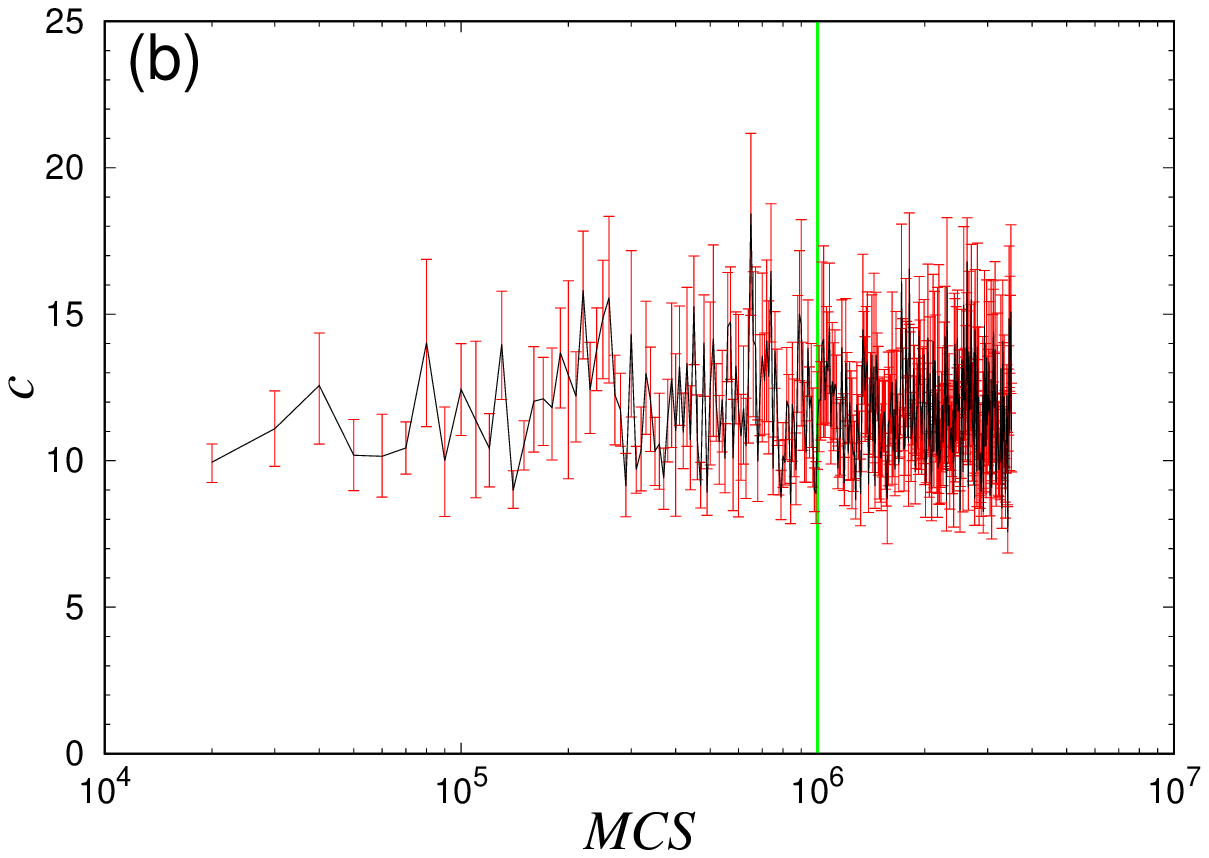}
    \includegraphics[width=40mm]{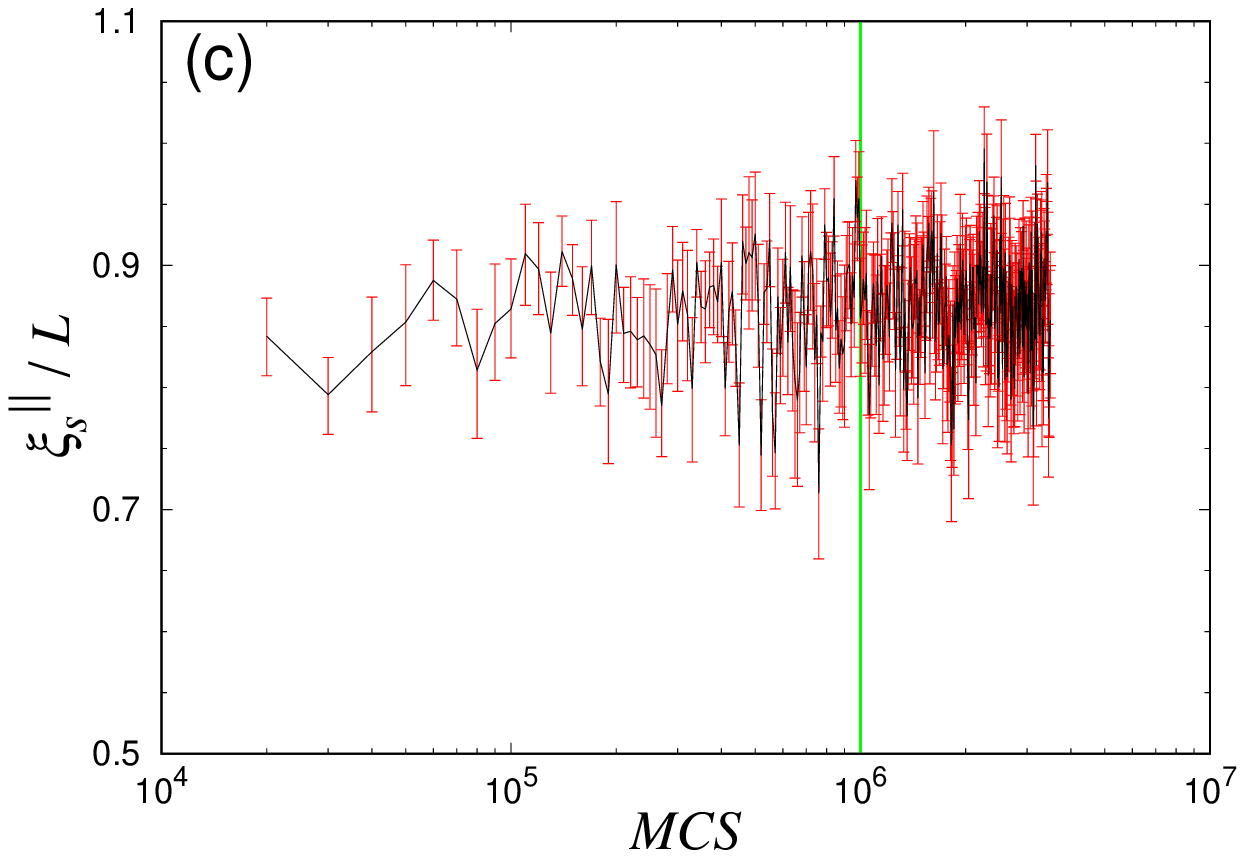}
    \includegraphics[width=40mm]{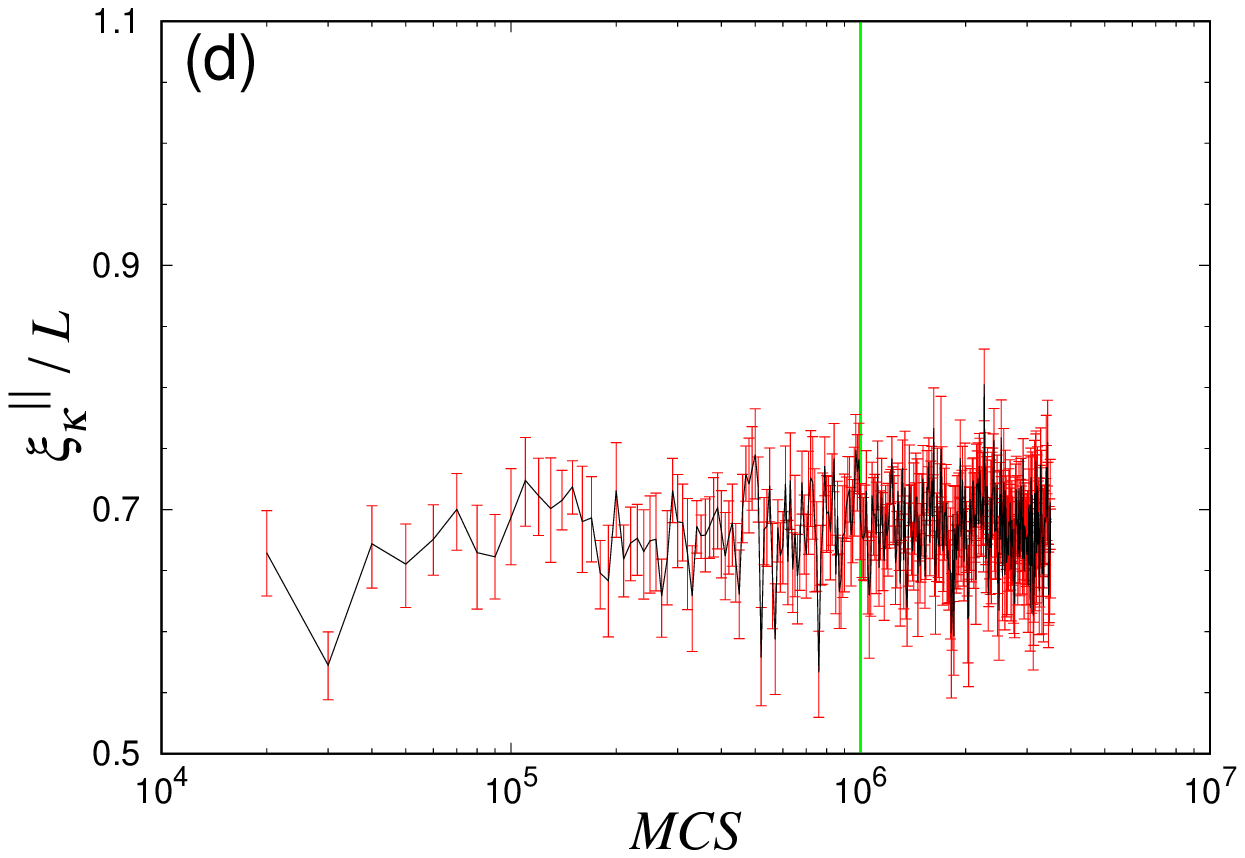}
   \caption{ 
(Color online) The MC-time dependence of  (a) the energy per spin $\bar e$, (b) the specific heat per spin $c$, (c) the intraplane spin correlation-length ratio $\xi_s^\parallel$, and (d) the intraplane chiral correlation-length ratio $\xi_\kappa^\parallel$. The lattice size is $L=384$. The temperature $T$ is set to the transition temperature $T=0.95727$.
} 
  \end{center}
\end{figure}
In order to be sure that MC simulations yield physical quantities {\it in thermal equilibrium\/}, the check of thermalization is important, especially for larger systems. In order to examine the thermalization, we monitor the MC-time dependence of various physical quantities to check that they reach stationary values. In Fig.1(a)-(d), we show the MC-time $t_{MC}$ dependence of (a) the energy per spin, (b) the specific heat per spin, (c) the intraplane spin correlation-length ratio and (d) the intraplane chiral correlation-length ratio on the logarithmic scale for our largest size $L=384$ taken at the transition temperature (to be determined below) $T=0.95727$. The short-time averaging of these observables are made over $10^4$ MCS at every $10^4$ MCS, and these short-time averaged values are plotted versus the elapsed MC time. As can be seen from the figures, all the quantities reach stationary values when $t_{MC}$ exceeds $\sim 10^6$ MCS. As mentioned, we discard first $10^6$ MCS for thermalization and use subsequent $2.5\times 10^6$ MCS to compute physical quantities. Measurements of physical quantities are made at every MCS.

\section{IV. The Monte Carlo data}

In this section, we present our MC data of the computed physical quantities in the transition region. The temperature ($T$) and size dependence of the energy is shown in Fig.2(a). While there develops a steep inflection-point anomaly for larger sizes, there is no appreciable discontinuity nor hysteresis indicative of a first-order transition. The temperature and size dependence of the specific heat is shown in Fig.2(b). There occurs a quite sharp divergent-like anomaly at $T\simeq 0.957$ signaling the occurrence of a thermodynamic phase transition. The size dependence of the peak height is shown in the inset. The peak height grows markedly with $L$, consistently with the previous works \cite{KawamuraMCH,KawamuraMCHXY}. 
% If the transition is of first-order, the peak height should grow proportional to $L^d=L^3$ ($d=3$). We shall return the issue of the order of the transition later in \S VI. 
%
\begin{figure}[t]
  \begin{center}
    \includegraphics[width=\hsize]{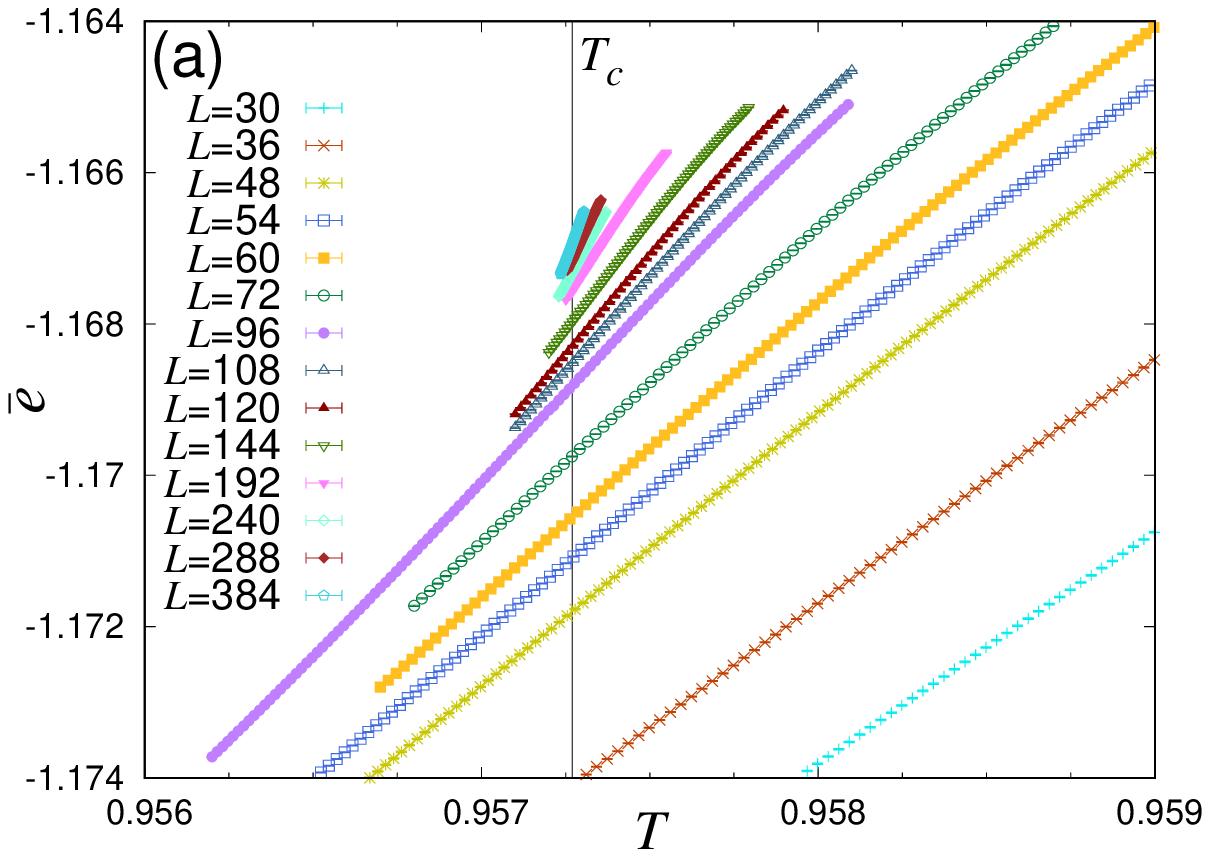}
    \includegraphics[width=\hsize]{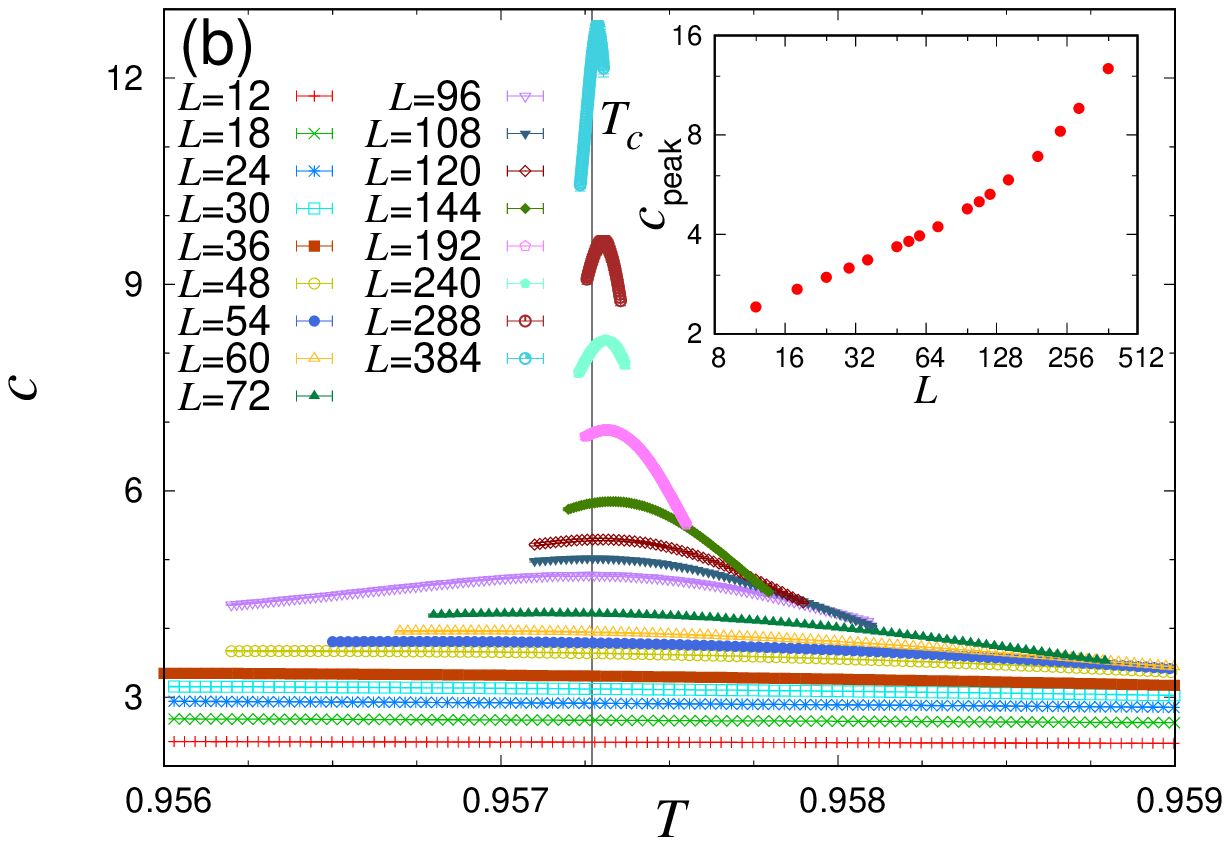}
   \caption{ 
(Color online) The temperature and size dependence of (a) the energy $\bar e$ and (b) the specific heat $c$, around the transition temperature $T_c$. The inset of (b) exhibits the size dependence of the specific-heat-peak height. 
} 
  \end{center}
\end{figure}

 The temperature and size dependence of the order parameter $m_{AF}$ is shown in Fig.3(a). With decreasing $T$ across $T_c$,  $m_{AF}$ exhibits a sharp rise signaling the onset of the AF LRO. No sign of hysteresis or discontinuity indicative of a first-order transition is observed again. In Fig.3(b), we show the temperature and size dependence of the intraplane spin correlation-length ratio $\xi_s^\parallel/L$. As $L$ is increased toward the thermodynamic limit $L\rightarrow \infty$, the correlation-length ratio $\xi/L$ should vanish from above at temperatures higher than $T_c$, approach unity from below at temperatures lower than $T_c$, and approach a nontrivial finite value just at $T=T_c$. Such a behavior in the  $L\rightarrow \infty$ limit entails that, for larger $L$, $\xi/L$ for different $L$ cross with each other with its crossing temperature converging to the bulk $T_c$. As expected, there occurs a crossing point between different size data in Fig.3(b), indicative of a magnetic phase transition.
\begin{figure}[t]
  \begin{center}
    \includegraphics[width=\hsize]{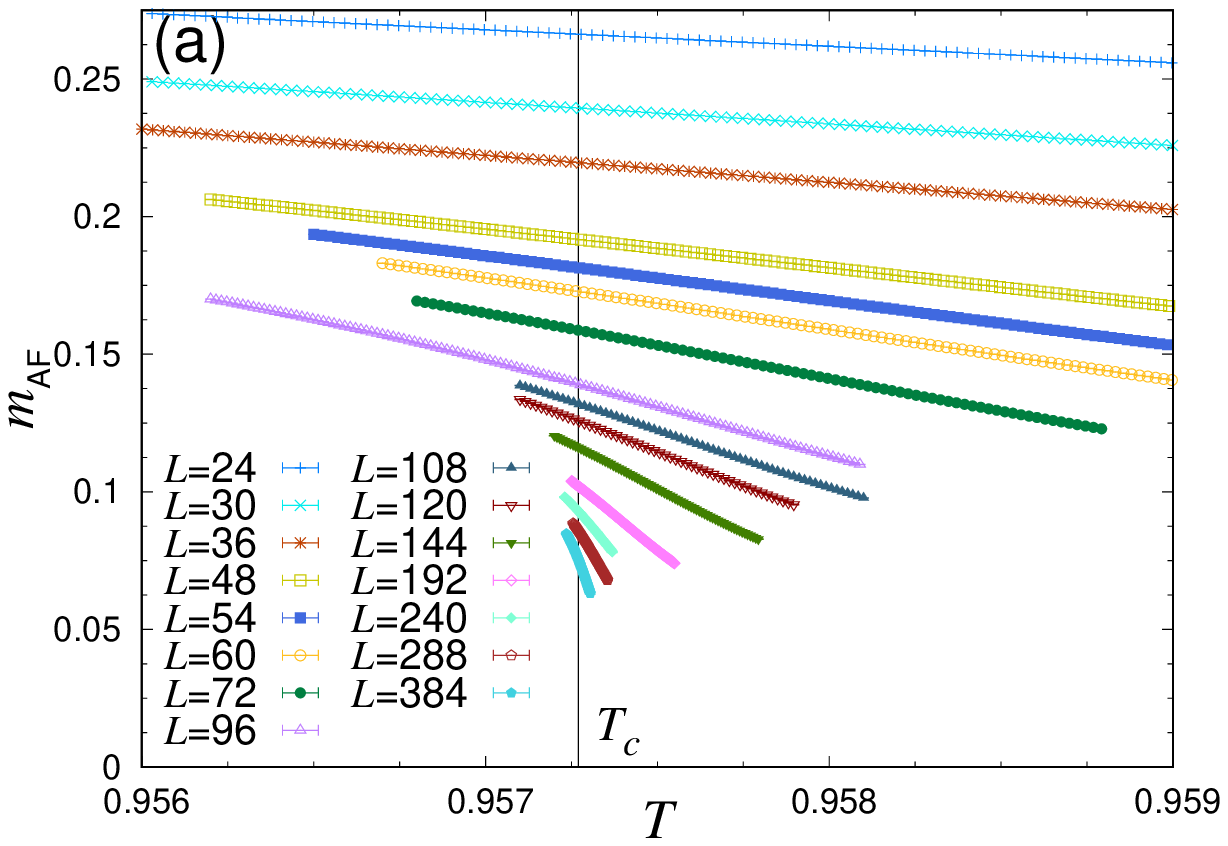}
    \includegraphics[width=\hsize]{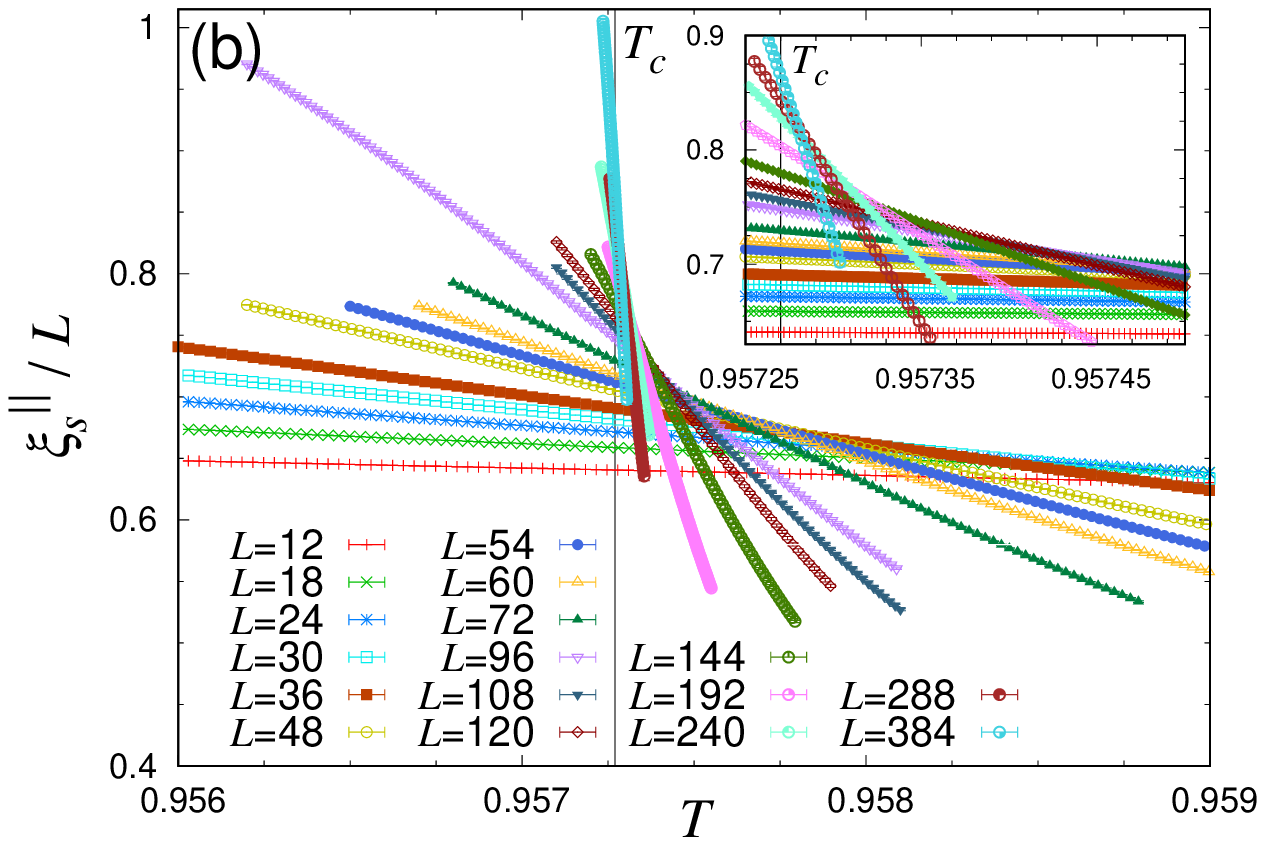}
   \caption{
(Color online) The temperature and size dependence of (a) the AF order parameter $m_{AF}$, and (b) the intraplane spin correlation-length ratio $\xi_s^\parallel/L$, around the transition temperature $T_c$. The inset of (b) is a magnified view of the transition region.
} 
  \end{center}
\end{figure}

 Similar plots are also given in Figs.4(a) and (b) for the chirality $\kappa$  and the associated intraplane chiral correlation-length ratio $\xi_\kappa^\parallel/L$. At almost the same temperature as that of the spin, the chirality also exhibits a sharp rise and the  chiral correlation-length ratio exhibits a crossing behavior. This observation strongly suggests that the spin and the chirality order at a common temperature, as was indicated by the previous works \cite{KawamuraMCH,KawamuraMCHXY}.

\begin{figure}[t]
  \begin{center}
    \includegraphics[width=\hsize]{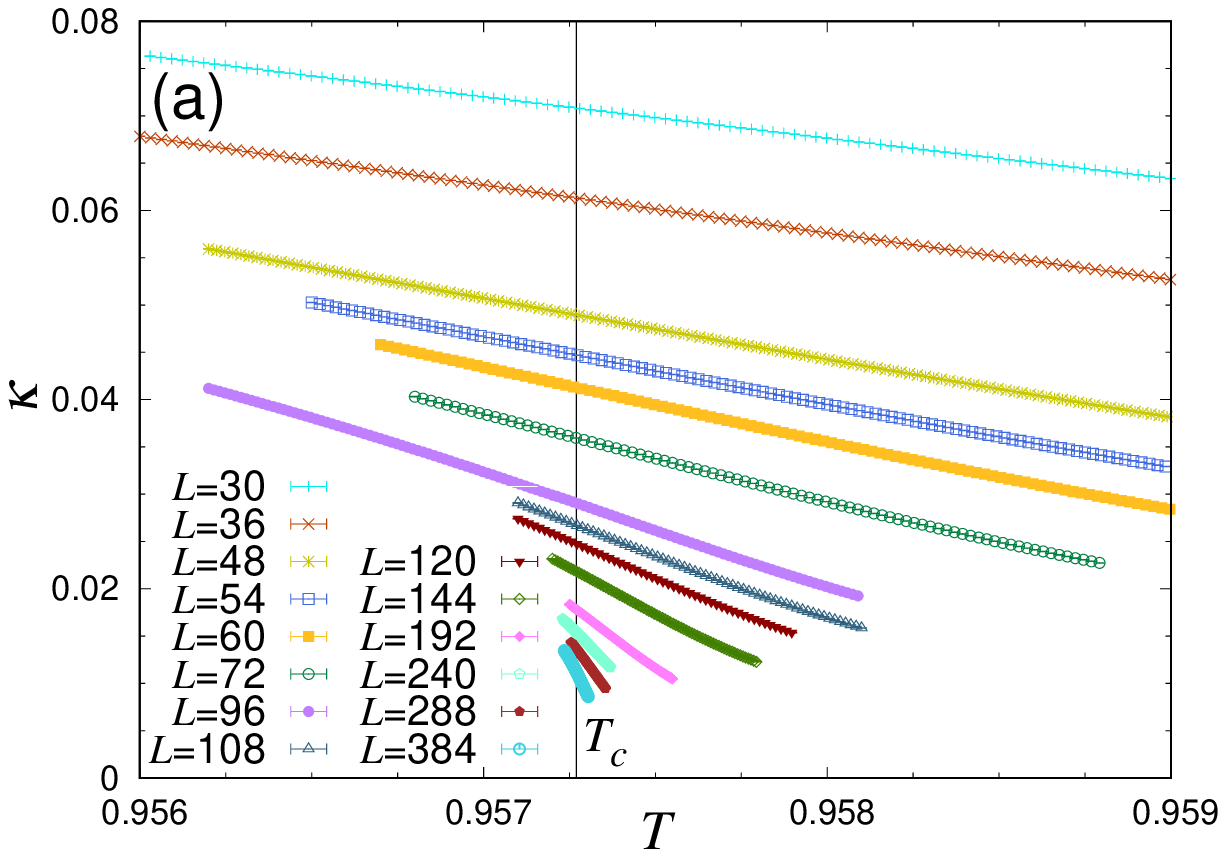}
    \includegraphics[width=\hsize]{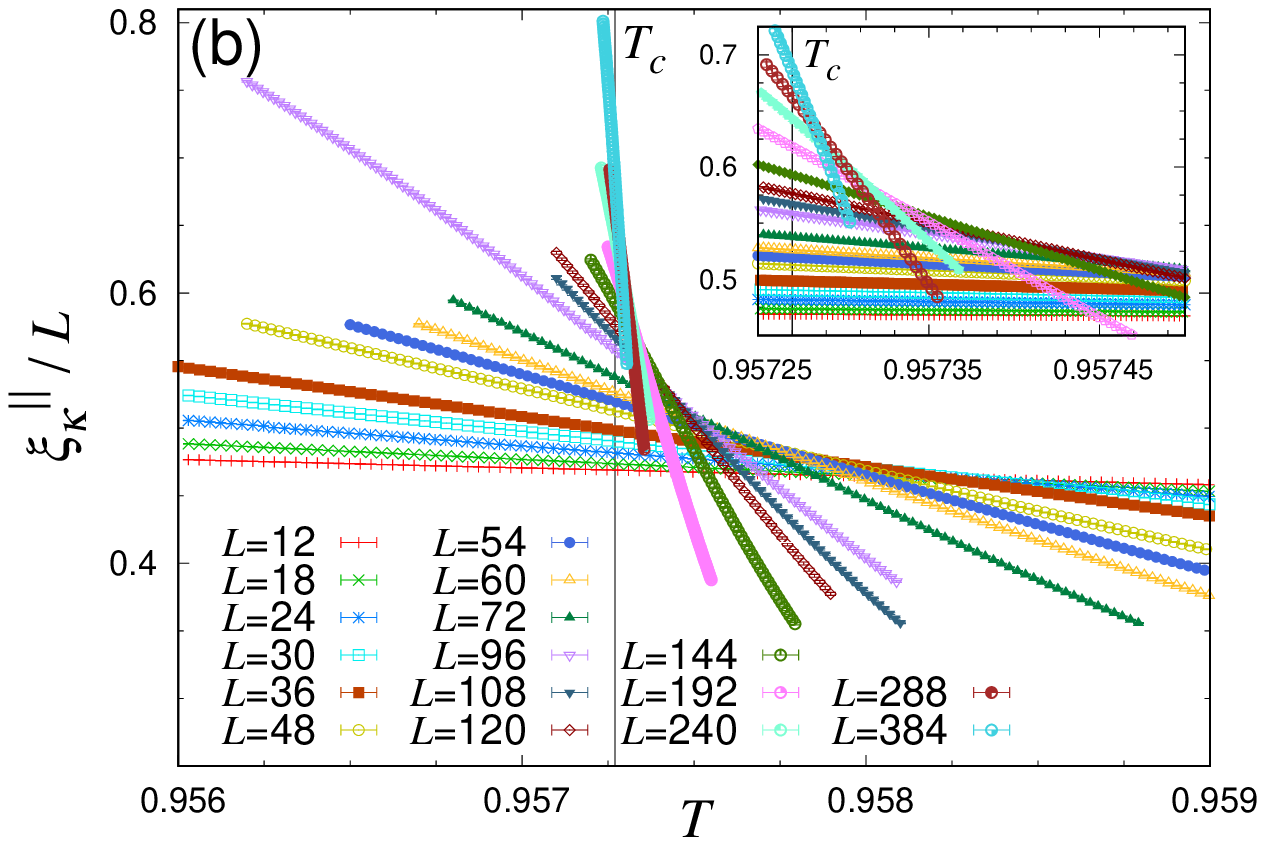}
   \caption{(Color online) The temperature and size dependence of (a) the chirality $\kappa$, and (b) the intraplane chiral correlation-length ratio $\xi_\kappa^\parallel/L$, around the transition temperature $T_c$. The inset of (b) is a magnified view of the transition region.
} 
  \end{center}
\end{figure}

The behaviors of some other quantities, including the energy Binder ration $g_e$ (Fig.S1), the $T$-derivative of the AF order parameter ${\rm d}m_{AF}/{\rm d}T$ (Fig.S2(a)), the interplane spin correlation-length $\xi_s^\perp /L$ (Fig.S2(b)), the spin Binder ratio $g_s$ (Fig.S3(a)), the $T$-derivative of the spin Binder ratio ${\rm d}g_{s}/{\rm d}T$ (Fig.S3(b)), and their chiral counterparts (Figs.4(a,b) and 5(a,b)) are given in Supplemental Material. The behaviors of all these computed quantities consistently suggest the occurrence of a single magnetic phase transition. The transition appears to be continuous, but we shall further examine this point later in \S VI.

\section{V. Determination of $T_c$}

In this section, on the basis of our numerical data for sizes as large as $L=384$, we wish to estimate the transition temperature $T_c$ as accurately as possible. Some physical quantities we compute exhibit a peak as a function of the temperature around $T_c$, which converges in the thermodynamic limit to the bulk $T_c$, and can be used in locating $T_c$. These quantities include the specific heat $c$, the energy Binder ratio $g_e$, the $T$-derivative of the AF order parameter ${\rm d}m_{AF}/{\rm d}T$,  the $T$-derivative of the spin Binder ratio ${\rm d}g_{s}/{\rm d}T$,  the $T$-derivative of the chirality ${\rm d}\kappa/{\rm d}T$, and the $T$-derivative of the chiral Binder ratio ${\rm d}g_{\kappa}/{\rm d}T$. In Fig.5, we plot the peak temperature $T_{{\rm peak}}(L)$ of these quantities versus the inverse lattice size $1/L$. As can be seen from the figure, many of $T_{{\rm peak}}(L)$ exhibit a monotonic size dependence, monotonically decreasing with increasing $L$ tending to a bulk transition temperature $T_c$ from above, whereas some others exhibit a non-monotonic size dependence: They first increase up to certain length scale $L_{{\rm cross}}$, then, for larger lattice sizes $L\gtrsim L_{{\rm cross}}$, they decrease tending to $T_c$. The crossover length scale is pretty long, $L_{{\rm cross}}\simeq 144$, signaling the existence of a rather large correction to the leading scaling. Anyway, the combined power-law fit of $T_{{\rm peak}}(L)$ for all the quantities at $L>L_{{\rm cross}}\simeq 144$ yields our first estimate of the bulk transition temperature, $T_c\simeq 0.95726-0.95729$. 
\begin{figure}[t]
  \begin{center}
    \includegraphics[width=\hsize]{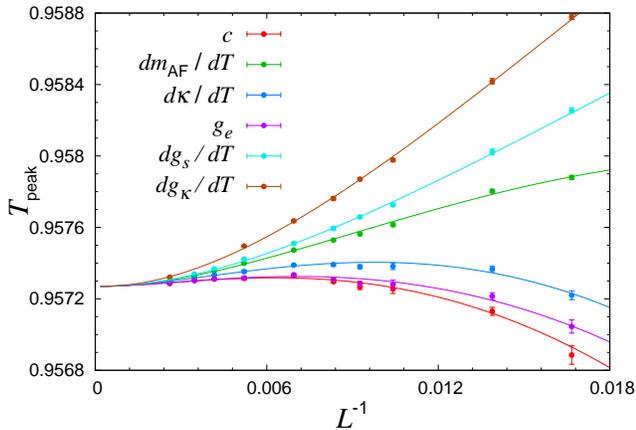}
   \caption{ 
(Color online) The peak temperatures $T_{{\rm peak}}$ of the specific heat $c$, the energy Binder ratio $g_e$, the $T$-derivative of the AF order parameter ${\rm d}m_{AF}/{\rm d}T$,  the $T$-derivative of the spin Binder ratio ${\rm d}g_{s}/{\rm d}T$,  the $T$-derivative of the chirality ${\rm d}\kappa/{\rm d}T$,  the $T$-derivative of the chiral Binder ratio ${\rm d}g_{\kappa}/{\rm d}T$ are plotted versus the inverse lattice size $1/L$. The fit is based on the finite-size scaling form eq.(\ref{Tpeak-fit}) with $T_c=0.95727$, $\nu=0.52$, $\omega_R=0.1$ and $\omega_I=0.7$: See \S VII for further details of the fit.
} 
  \end{center}
\end{figure}

 In order to get a more precise estimate of $T_c$, we employ the spin  correlation-length ratios, $\xi_s^\parallel/L$ and $\xi_s^\perp/L$. As these quantities are dimensionless, their size dependence is insensitive to the correlation-length exponent $\nu$, depending only on the correction-to-scaling exponent $\omega$. As mentioned above, in the thermodynamic limit $L\rightarrow \infty$, the correlation-length ratio $\xi/L$ goes to zero at $T>T_c$ and to infinity at $T<T_c$. Just at $T=T_c$, it goes to a finite value $(\xi/L)^*$ as 
\begin{equation}
\xi_L\approx (\xi/L)^*(1+aL^{-\omega}), 
\label{xi-fit}
\end{equation}
where $a$ is a nonuniversal constant. In Fig.6(a) and (b), we plot the (a) intraplane and (b) interplane spin correlation-length ratios as a function of $1/L$ for several temperatures in the transition region, and try to fit the data by the finite-size-scaling form given in eq.(\ref{xi-fit}). Overall, as can be seen from Fig.6, $\xi_s/L$ tends to increase as the system size $L$ is increased, whereas a closer look of the data reveals a systematic changeover occurring. At the lower temperature $T=0.957260$, both correlation-length ratios $\xi_s^\parallel$ and $\xi_s^\perp$ exhibit a sharp increase toward $1/L\rightarrow 0$, yielding the fitted correction-to-scaling exponent $\omega$ close to zero, say, $\omega\simeq 0.0032$. This indicates that the temperature $T=0.957260$ is actually lower than $T_c$. By contrast, at the higher temperature $T=0.957280$, some of the  correlation-length ratios begin to decrease for the largest size, indicating that the temperature $T=0.957280$ is actually located above $T_c$. From such a changeover seen in Figs.6(a) and (b), we estimate $T_c=0.957270\pm 0.000004$. The estimated $T_c$ is consistent with the earlier MC estimates on the same model within the quoted error bars, {\it i.e.\/}, $T_c=0.958(4)$ \cite{KawamuraMCHXY}, $T_c=0.9576(2)$ \cite{Bhattacharya}, and $T_c=0.9577(2)$ \cite{Mailhot}, but orders of magnitudes more precise.
\begin{figure}[t]
  \begin{center}
    \includegraphics[width=\hsize]{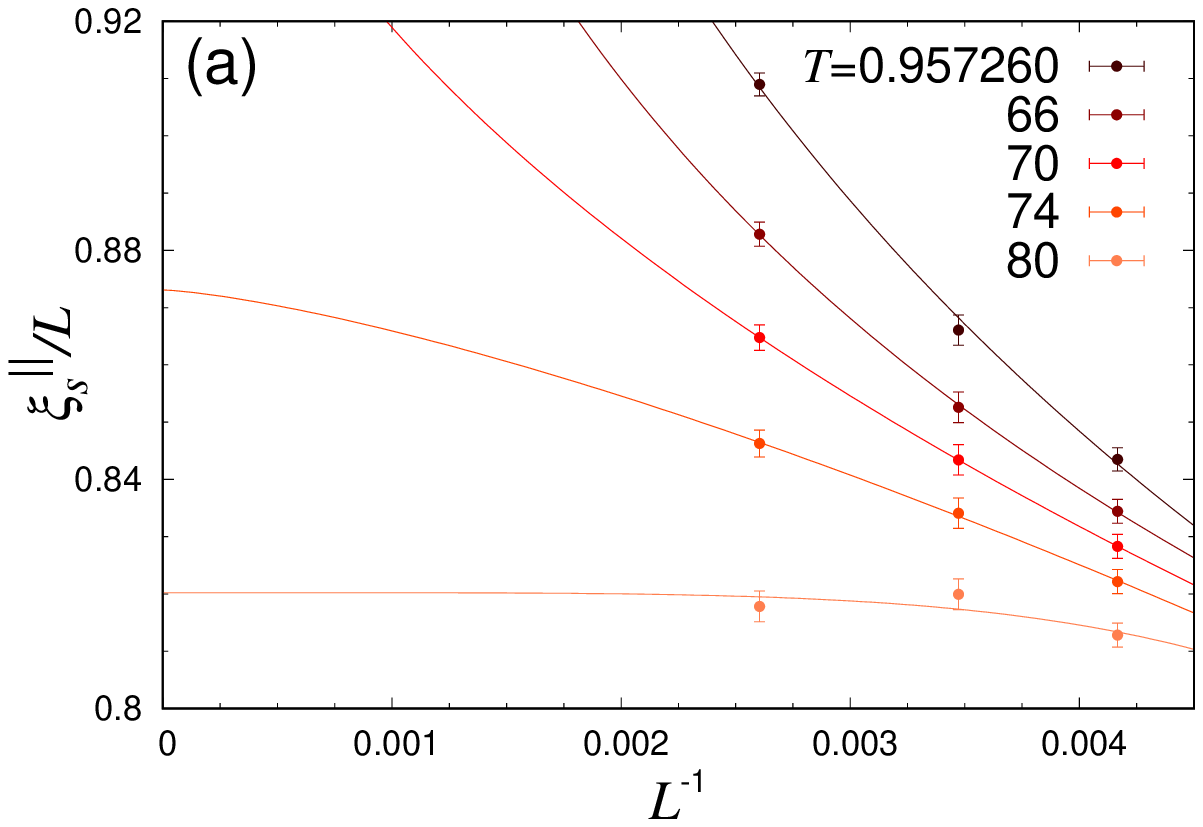}
    \includegraphics[width=\hsize]{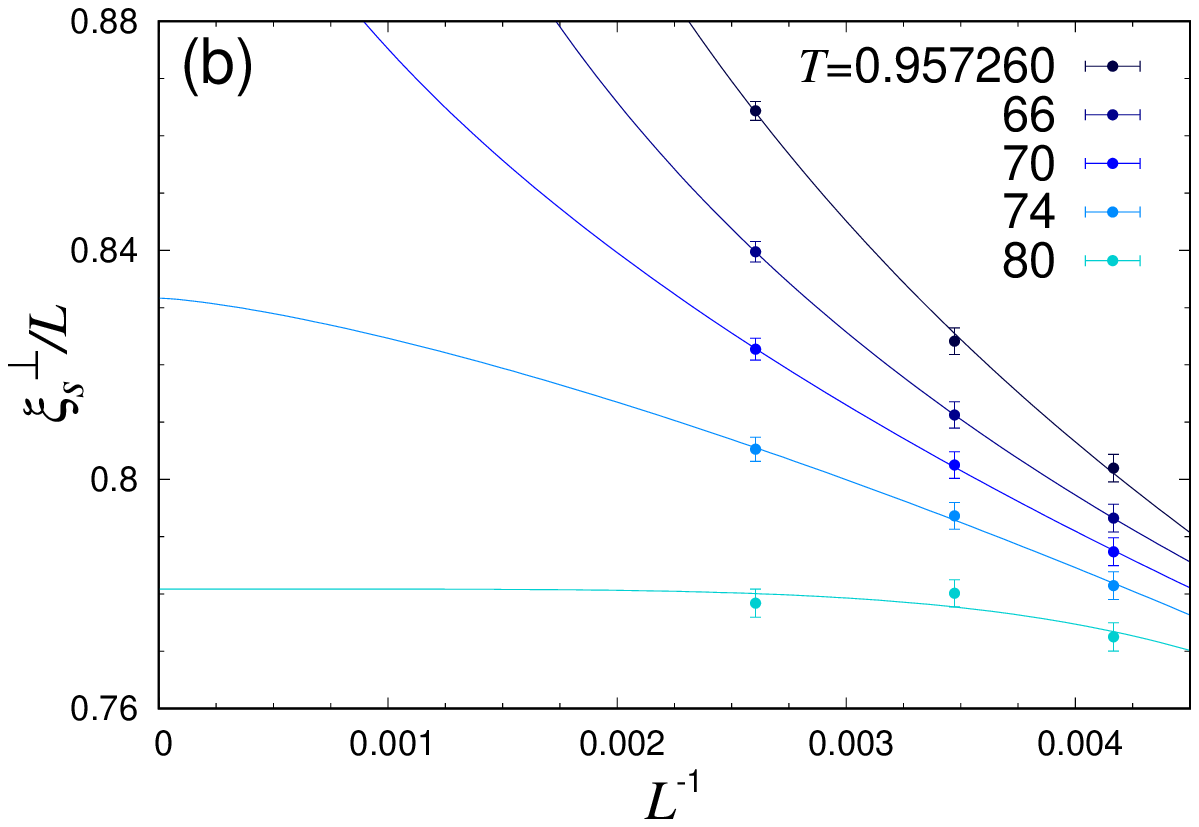}
   \caption{ 
(Color online) (a) The intraplane spin correlation-length ratio $\xi_s^\parallel /L$ and (b) the interplane spin correlation-length ratio $\xi_s^\perp /L$ are plotted versus the inverse lattice size $1/L$ at several temperature around $T_c=0.957270$. The fit is based on eq.(\ref{xi-fit}) with using the data of $L=240$, 288 and 384.
} 
  \end{center}
\end{figure}

 In determining $T_c$, one sometimes employs the crossing temperatures $T_{{\rm cross}}$ of the dimensionless quantities, {\it e.g.\/}, the correlation-length ratio and the Binder ratio, of two different sizes, $L$ and $sL$ ($s>1$). In systems exhibiting a finite-$T$ transition, these dimensionless quantities of two different sizes often cross at a size-dependent temperature $T_{{\rm cross}}$, which converges to the bulk $T_c$ in the infinite-size limit. We also perform here such an analysis to estimate $T_c$. Some of the details of the analysis are given in Supplemental Materials. In fact, in the presence of the nontrivial and significant correction to scaling as in the present case, the extrapolation of $T_{{\rm cross}}$, which are defined for the two different sizes and are more susceptible to the correction-to-scaling, might behave worse than that of $T_{{\rm peak}}$ defined for the single size. Nevertheless, we find that the extrapolated $T_c$ is basically consistent with the one obtained from $T_{{\rm peak}}$ and the correlation-length ratio as quoted above: See Supplemental Materials for further details.

\section{VI. The order of the transition}

 In this section, we wish to examine  the order of the magnetic transition of the model, which has remained controversial for years. In Fig,7, we show the energy distribution $P(e)$ of the model around $T_c=0.95727$ for larger lattices of $L=144$, 288 and 384, to examine whether $P(e)$ exhibits a single-peak characteristic of a continuous transition or double peaks characteristic of a first-order transition. (Of course, the occurrence of the double-peak structure in $P(e)$ for finite $L$ does not necessarily mean a first-order transition. One needs to check carefully that such a double-peak structure persists in the $L\rightarrow \infty$ limit.) As can be seen from Fig.7, $P(e)$ exhibits a single peak for all sizes and at any temperature. Any sign of the double-peak structure signaling a first-order transition is not detected for all the sizes and temperatures studied, even including the ones not explicitly shown in Fig.7. Hence, we conclude that the transition of the model is continuous. In Ref.\cite{Diep}, by observing the double peaks in the energy distribution for $L=120$ and 150 by means of the Wang-Landau method \cite{WangLandau}, Ngo and Diep concluded that the transition of the model was actually first-order, arguing that the system sizes studied in the previous MC simulations on the same model were too small. However, our largest size $L=384$ is considerably larger than the largest size studied in Ref.\cite{Diep}, $L=150$. Yet, we do not observe any sign of the double-peak structure reported in Ref.\cite{Diep}. We confirm that even for $L=144$ and $T=0.957240$ (quite close to $L=150$ and $T=0.957242$ studied in Ref.\cite{Diep}) the distribution is definitely single-peaked as shown in Fig.7, contrary to the report of Ref.\cite{Diep}. Note that our energy resolution (the width of the bin of $P(e)$), $2.5\times 10^{-5}$ for $L=144$ and $1.5\times 10^{-6}$ for $L=288$, 384, is much better than the latent heat reported in Ref.\cite{Diep}, 0.0025, so that we cannot miss the double-peak structure at the level reported in Ref.\cite{Diep} if it really exists. % Hence, we suspect that there might have been something wrong in Ref.\cite{Diep} in the application of the Wang-Landau method.
\begin{figure}
  \begin{center}
    \includegraphics[width=\hsize]{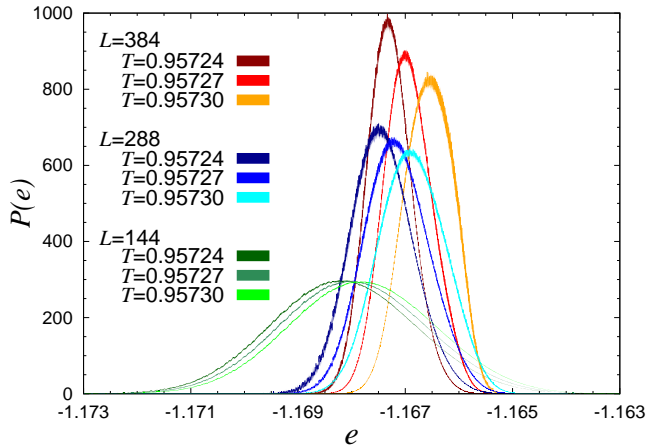}
   \caption{
(Color online) The energy distribution $P(e)$ around the transition temperature $T=T_c=0.95727$ for the sizes $L=384$, 288 and 144.
} 
  \end{center}
\end{figure}

 Another evidence of the continuous nature of the transition comes from the energy Binder ratio $g_e$, which exhibits a single peak as a function of the temperature around $T=T_c$ as shown in Fig.S1. In the thermodynamic limit, $g_e$ at $T=T_c$ should take a value equal to unity if the transition is continuous, while it should take a value greater than unity if the transition is of first-order \cite{Binder}. Thus, in Fig.7(a), we plot $g_e-1$ at the transition temperature $T=T_c=0.95727$ versus the inverse lattice size $1/L$. For a first-order transition, $g_e-1$ should exhibit a size-scaling of the form $g_e-1=\delta_g +\frac{1}{L^3}+\cdots $ with $\delta_g >0$. As can be seen from the figure, $g_e -1$ becomes quite small for our largest lattice size $L=384$. We perform a simple power-law fit of the data to the form  $g_e=g_e^\infty+cL^{-\theta}$ in the $L$-range of $L_{{\rm min}}\leq L\leq L_{{\rm max}}=384$, and the resulting $g_e^\infty$ and $\theta$ are given in the inset of Fig.7(a) as a function of $L_{{\rm min}}$. The extrapolated value of $g_e^\infty$ is already as small as $10^{-7}$ for larger $L_{{\rm min}}$, and tends to decrease further on increasing $L_{{\rm min}}$. As can be seen from the inset, the effective exponent $\theta$ describing the $L$-dependence of $g_e-1$ tends to decrease from three, further deviating from the value of the first-order transition. In Fig.7(b), we replot the same data as of Fig.7(a) versus $1/L^3$. As can be seen from the figure, the data for larger $L$ (smaller $1/L^3$) exhibits {\it a decrease  stronger than $1/L^3$ toward zero}, deviating from the finite-size scaling form expected for a first-order transition. Together with the single-peaked energy distribution, the observed behaviors provide a strong support of the continuous nature of the transition.

\begin{figure}
  \begin{center}
    \includegraphics[width=\hsize]{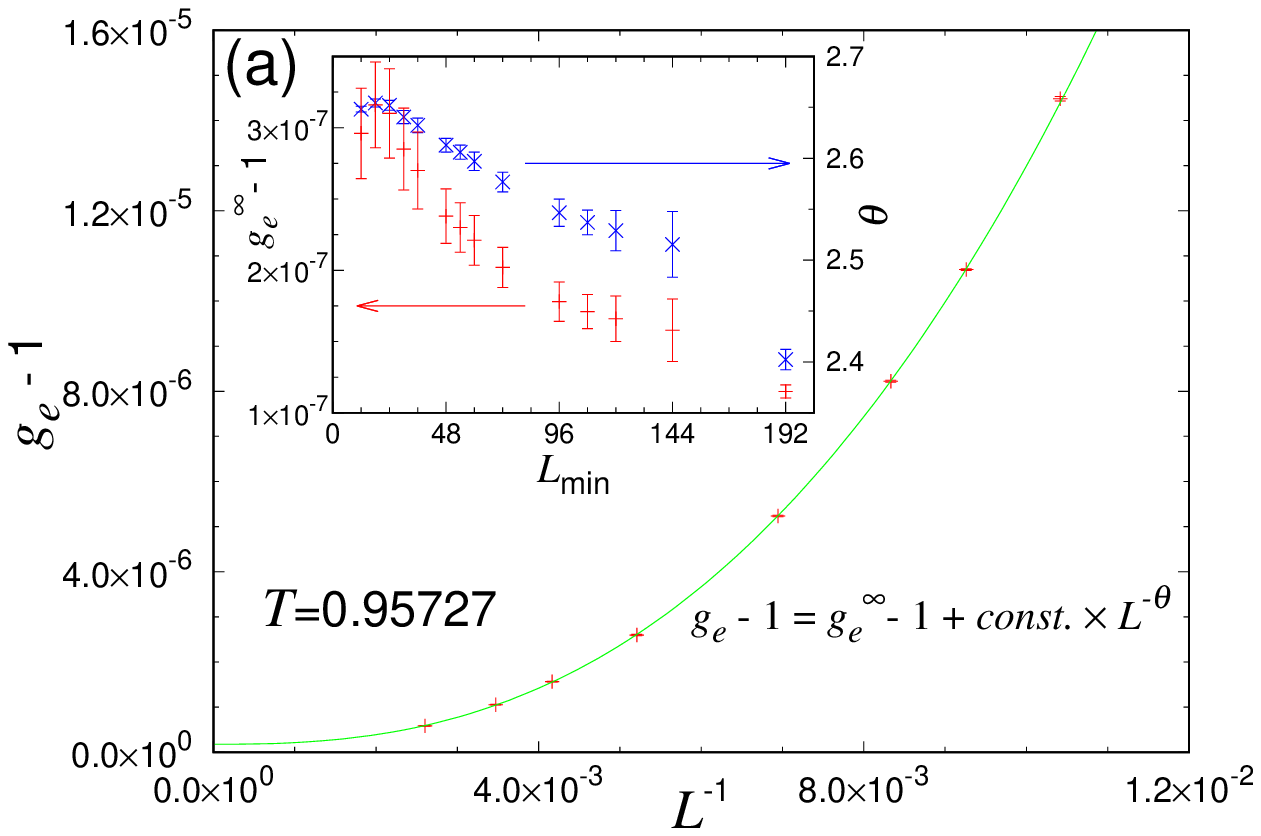}
    \includegraphics[width=\hsize]{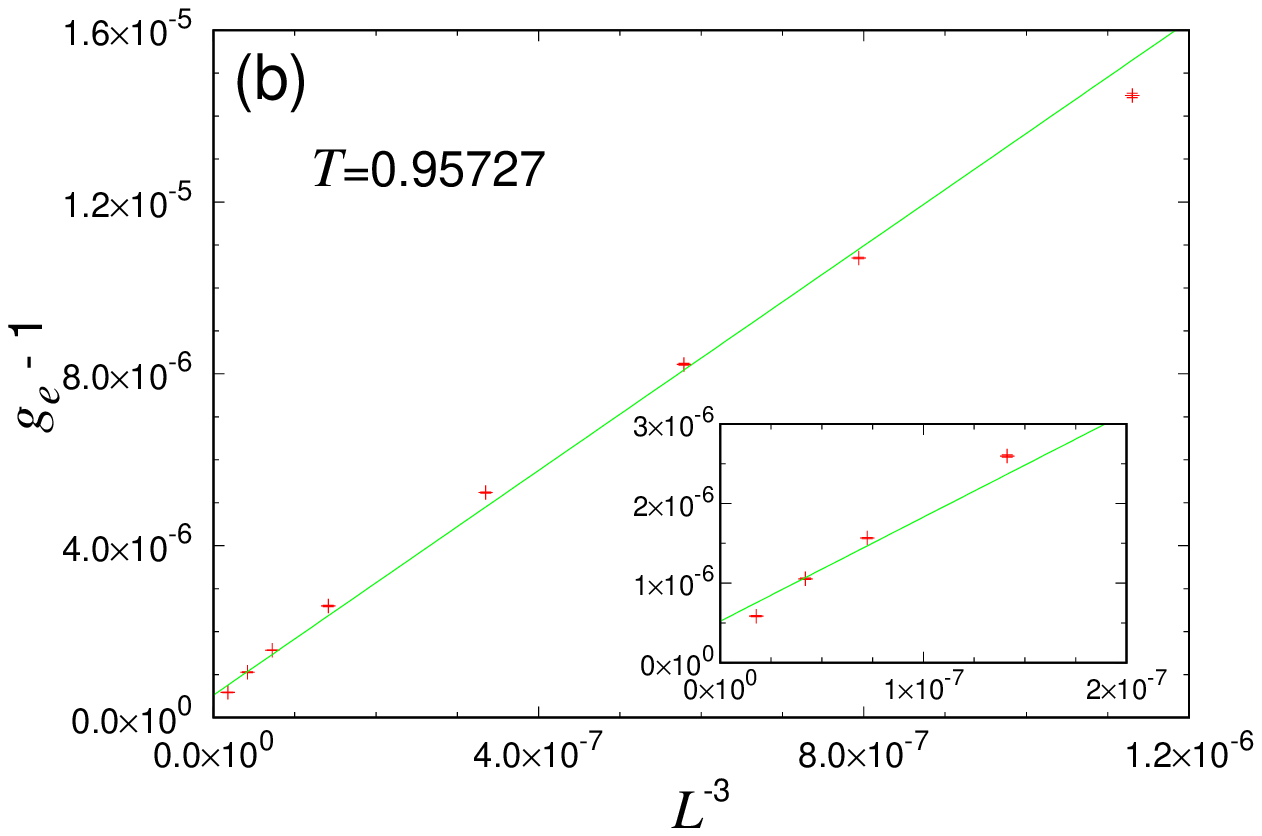}
   \caption{(Color online) (a) The energy Binder ration $g_e$ at the transition temperature $T=T_c=0.95727$ are plotted versus the inverse lattice size $1/L$. The inset represents the extrapolated value of $g_e^\infty$ and $\theta$ based on the extrapolation formula $g_e=g_e^\infty+{\rm const.}\times L^{-\theta}$. (b) The $g_e$ data are replotted versus $1/L^3$. In both (a) and (b), the fitting curves employ the data in the range $96 \leq L \leq 384$; (a) the power-law fit, and (b) the linear fit.
} 
  \end{center}
\end{figure}

\section{VII. Analysis of the critical properties}

 After establishing the continuous nature of the transition, we now wish to investigate its critical properties, {\it i.e.\/}, determine various critical exponents on the basis of our precise estimate of the transition temperature, $T_c=0.957270\pm 0.000004$. Since similar analysis in \S IV has already indicated that there exists a large correction to the leading scaling, care has to be taken.

\subsection{1. Analysis without the correction} 

In this subsection, we wish to examine the critical behavior of the model by employing the leading term only, without explicitly invoking the correction term. The exponent arising from such an analysis would only be an {\it effective\/} exponent, rather than the true asymptotic exponent. Still, the analysis would be informative giving some information about the  correction.

\begin{figure}
  \begin{center}
    \includegraphics[width=\hsize]{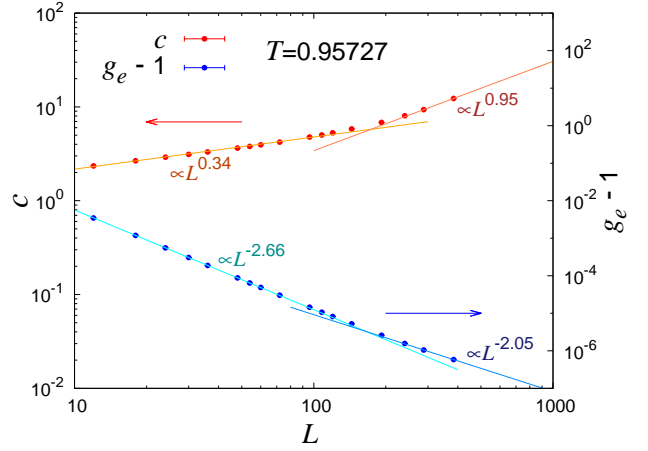}
   \caption{
(Color online) The specific heat $c$ and the energy Binder ratio $g_e$ at $T=T_c=0.95727$ are plotted versus the system size $L$ on the double-logarithmic scale.
} 
  \end{center}
\end{figure}

 In Fig.9, we show the size dependence of the specific heat $c$ and of the energy Binder ratio $g_e$ at $T=T_c=0.95727$ on the double-logarithmic plot. The expected leading-scaling forms for $\alpha>0$ should be
\begin{eqnarray}
c &\sim& L^{\alpha/\nu}=L^{\frac{2}{\nu}-3}, 
\label{c}
\\
g_e-1 &\sim& = L^{-(3-\alpha/\nu)} = L^{-2(3-1/\nu)}, \ \ \alpha>0,
\label{ge}
\end{eqnarray}
where we have employed the hyperscaling relation $\alpha=2-d\nu=2-3\nu$. As we could not find in the literature the expression of the relevant exponent $\theta$ for the energy Binder ratio $g_e-1 \approx L^{-\theta}$, we show its derivation in Appendix. We have $\theta=3-\alpha/\nu=2(3-1/\nu)$. (The corresponding expression for general dimension $d$ is given in Appendix.)

 The asymptotic size dependence of $c$ and of $g_e$ is described by the exponent $\nu$. As can be seen from Fig.9, the data exhibit continuously-varying slopes versus $L$, and cannot be fitted by a single straight line. For $c$, the slope changes from 0.34 describing the smaller-size data of $12 \leq L\leq 144$, to 0.95 describing the larger-size data of $288\leq L\leq 384$, while, for $g_e$, it changes from -2.66 to -2.05. Thus, the effective exponent varies considerably depending on the size $L$, indicating the existence of the large correction to the leading scaling. 
\begin{figure}
  \begin{center}
    \includegraphics[width=\hsize]{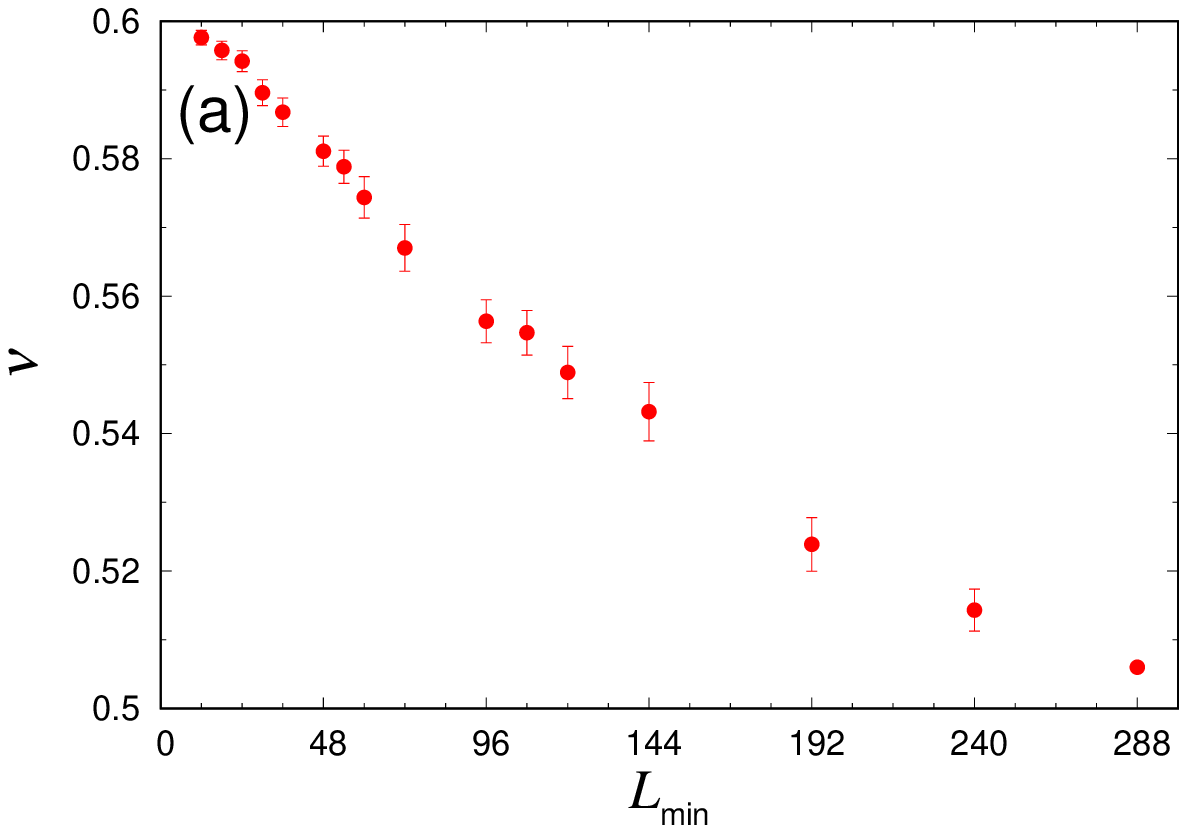}
    \includegraphics[width=\hsize]{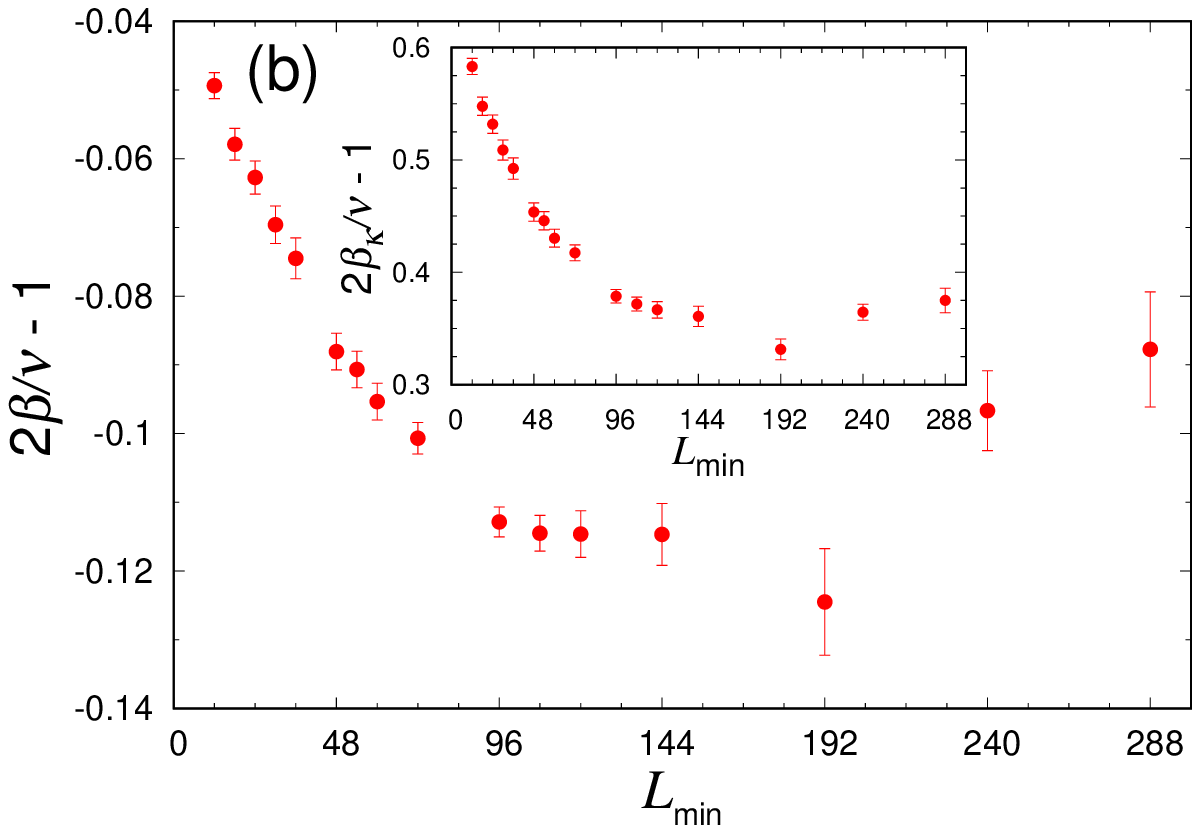}
   \caption{
(Color online) The effective exponents $\nu$ and $2\beta/\nu-1=\eta$ obtained by the scaling fit without the correction term, eqs. (21)-(26), are plotted versus $L_{{\rm min}}$, the minimum lattice size used in the fit. The inset of (b) is the corresponding plot for the effective chirality exponent $2\beta_\kappa/\nu-1$.
} 
  \end{center}
\end{figure}

 Similar behaviors of continuously-varying effective exponents are observed, though to less extent, in other quantities as well, including the AF order parameter $m_{AF}$, the AF susceptibility $\chi_{AF}$, the chirality $\kappa$, and the chiral susceptibility $\chi_\kappa$. The asympotic behaviors of $m_{AF}$ and $\chi_{AF}$ are described by the order-parameter exponent $\beta$, the ordering susceptibility exponent $\gamma$, and the critical-point-decay exponent $\eta$ as
\begin{eqnarray}
m_{AF} &\sim& L^{-\frac{\beta}{\nu}} = L^{-\frac{1+\eta}{2}}, 
\label{mAF}
\\
\chi_{AF} &\sim&  L^{\frac{\gamma}{\nu}} = L^{2-\eta},
\label{chiAF}
\end{eqnarray}
while those of $\kappa$ and $\chi_\kappa$ are described by the chirality exponent $\beta_\kappa/\nu$ as
\begin{eqnarray}
\kappa &\sim&  L^{-\frac{\beta_\kappa}{\nu}},
\label{kappa}
\\
\chi_\kappa \sim L^{\gamma_\kappa/\nu} &=& L^{(d-2\frac{\beta_\kappa}{\nu})} = L^{3-2\frac{\beta_\kappa}{\nu}}
\label{chikappa}
\end{eqnarray}

 In order to extract more quantitative information about the effective exponents, we fit the data by the above scaling forms in the size range of $L_{{\rm min}}\leq L \leq L_{{\rm max}}\equiv 384$ and extract the effective exponent as a function of $L_{{\rm min}}$ (the maximum size is fixed to $L_{{\rm max}}=384$). In order to estimate the effective exponents, we employ the combined fit of $c$ and $g_e-1$ for $\nu$, that of $m_{AF}$ and $\chi_{AF}$ for $\frac{2\beta}{\nu}-1=\eta$, and that of $\kappa$ and $\chi_\kappa$ for $\frac{2\beta_\kappa}{\nu}-1$. The results are shown in Fig.10(a) for $\nu$, in Fig.10(b) for $\frac{2\beta}{\nu}-1=\eta$, and in Fig.10(c) for $\frac{2\beta_\kappa}{\nu}-1$. As can be seen from Fig.10(a), the exponent $\nu$ tends to get smaller as $L_{{\rm min}}$ increases. By contrast, the exponents $\frac{2\beta}{\nu}-1=\eta$ and $\frac{2\beta_\kappa}{\nu}-1$ show a non-monotonic behavior as a function of $L_{{\rm min}}$. With increasing $L_{{\rm min}}$, they decrease for smaller $L_{{\rm min}}$, but exhibits a turnover and increase for larger $L_{{\rm min}}$. Concerning $\eta$, it changes from small positive numbers to near-zero or even small negative numbers for smaller $L_{{\rm min}}$, and exhibits a turnover toward small positive numbers at larger $L_{{\rm min}}$. In  any case, the observed significant size dependence of the effective exponents on the system size $L_{{\rm min}}$ warrant the inclusion of appropriate correction terms into the finite-size scaling analysis, which we try in the following subsections.

\subsection{2. Analysis with real correction exponents} 

First, we try to include a single correction term with the correction-to-leading-scaling exponent $\omega$ in the from of a simple multiplicative factor $(1+aL^{-\omega})$. In fact, however, we find that the inclusion of a single correction term does not much improve the fit. This inadequacy might be seen from the non-monotonic behavior of the effective exponents shown in Fig.10(b). Namely, the correction term of the form  $(1+aL^{-\omega})$ can describe only the monotonic change of the effective exponent, but not the non-monotonic one. To describe the non-monotonic behavior, one needs at least {\it two\/} corrections terms, {\it i.e.\/}, the one with distinct exponents, $\omega_1$ and $\omega_2$.

\begin{figure}
  \begin{center}
    \includegraphics[width=\hsize]{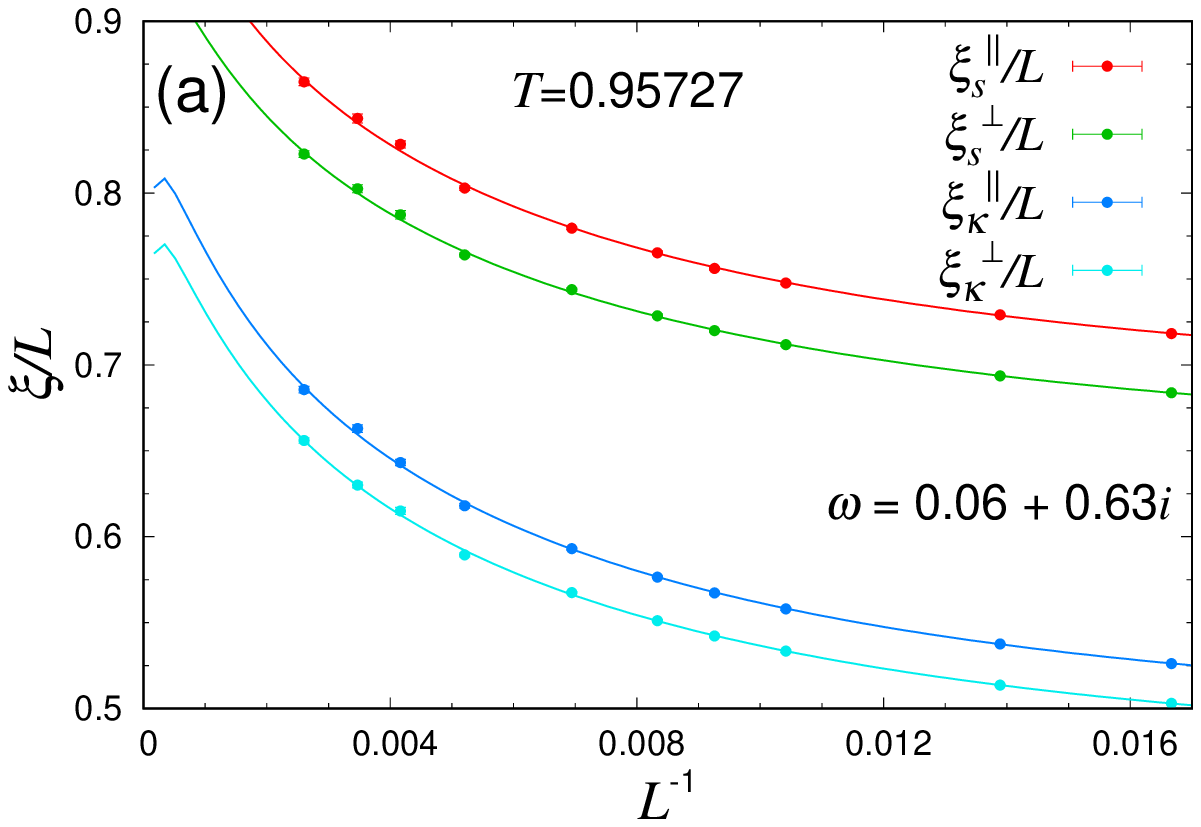}
    \includegraphics[width=\hsize]{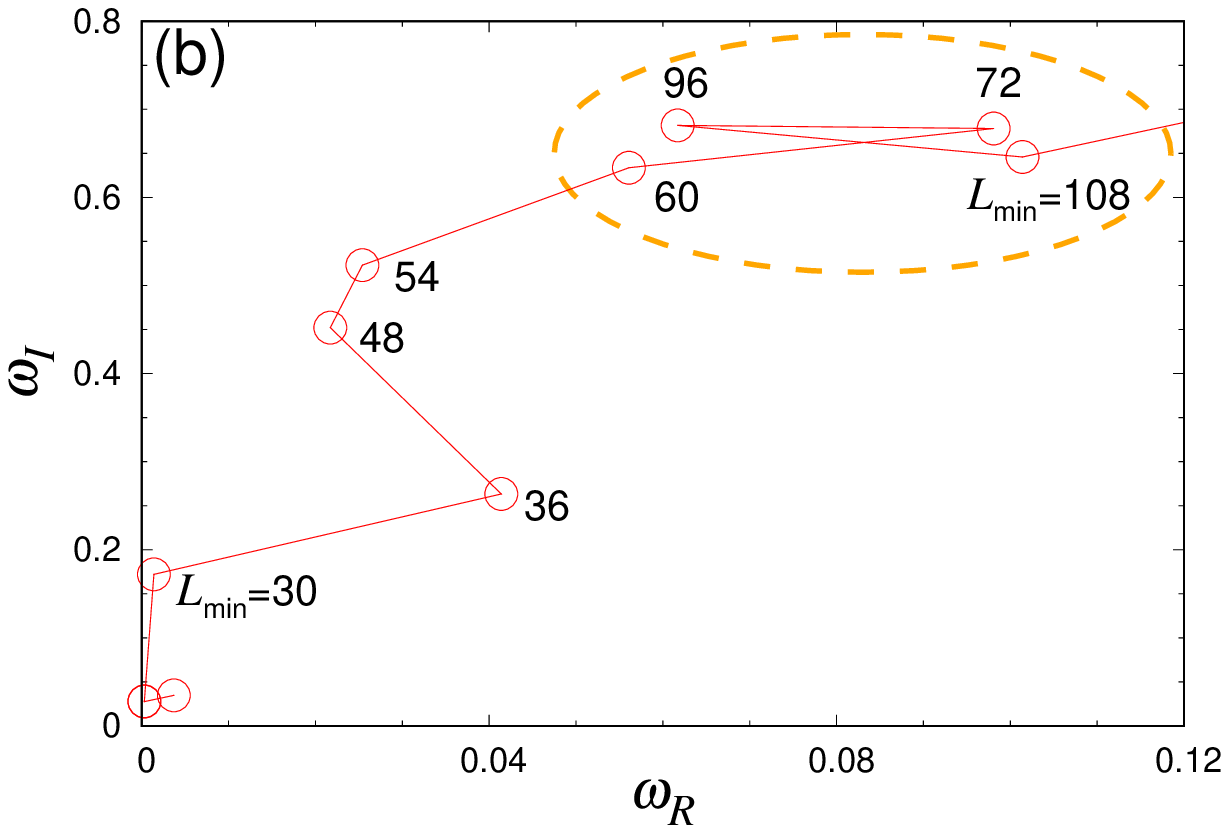}
   \caption{
(Color online) (a) The scaling plot of the spin and chiral correlation-length ratios at the transition temperature $T=T_c=0.95727$ both for the intra- and inter-plane correlations. The scaling employed includes the correction term with a complex correction-to-scaling exponent, eq.(\ref{complex-fit}), with $L_{{\rm min}}=60$. (b) The optimal value of the complex-valued correction-to-scaling exponent $\omega=\omega_R+i\omega_I$ are plotted in the ($\omega_R,\omega_I$)-plane for various choices of $L_{{\rm min}}$, the minimum lattice sized used in the fit.
} 
  \end{center}
\end{figure}
In this and following subsections, we perform the finite-size-scaling analysis by including {\it two\/} correction terms with the two correction-to-leading-scaling exponents $\omega_1$ and $\omega_2$. The standard way might be to assume two positive exponents, $0 < \omega_1 < \omega_2$. Namely, for the physical quantity $X$ at $T=T_c$, we assume the size-scaling form
\begin{equation}
X \approx  L^x(1+a_1L^{-\omega_1} + a_2L^{-\omega_2}),
\label{tworeal-fit}
\end{equation}
where $x$ is an appropriate critical exponent, $a_1$ and $a_2$ being nonuniversal coefficients. 

Higher-order perturbative RG analysis suggested that the appropriate FP might be of the ``focus''-type with a {\it complex-valued\/} correction-to-scaling exponent $\omega=\omega_R+i\omega_I$ ($\omega_R>0$) \cite{Calabrese1,Calabrese2}. Thus, we shall also examine in the next subsection the correction term described by a single complex correction-to-scaling exponent, which of course contains two real exponents $\omega_R$ and $\omega_I$. In this subsection, we first examine the standard correction terms containing two real exponents $\omega_1$ and $\omega_2$ as described by the scaling form eq.(\ref{tworeal-fit}) above. 

 As eq.(\ref{tworeal-fit}) has many fitting parameters, it turns out that the fitting usually leads to many local minima with comparable $\chi^2$-values. Hence, one needs to be careful not to miss the true minima with the optimal $\chi^2$-value. We begin our analysis with the correlation-length ratios. Since these quantities are dimensionless, the exponent $x$ does not appear in the scaling form eq.(\ref{tworeal-fit}) so that one can concentrate on the correction-to-scaling exponents. In Fig.S6(a), we show all the local minima obtained by our fitting of the correlation-length ratios in the $\omega_1$ vs. $\omega_2$ plane for the case of $L_{{\rm min}}=60$, where the color of the data points represents the associated $\chi^2$-value. We perform the combined fit for $\xi_s^{\parallel}/L$, $\xi_s^{\perp}/L$, $\xi_\kappa^{\parallel}/L$ and $\xi_\kappa^{\perp}/L$ for various values of $L_{{\rm min}}$. As can be seen from Fig.S6(a), there indeed exist many local minima in the fit. The best fit is obtained at $\omega_1\simeq 0.40$ and $\omega_2\simeq 0.45$, and the resulting fitting curves of each $\xi_s^{\parallel,\perp}/L$ and  $\xi_\kappa^{\parallel,\perp}/L$ are shown in Fig.S6(b). If $L_{{\rm min}}$ is varied, the resulting best values of $\omega_1$ and $\omega_2$ vary somewhat. How these best values of $\omega_1$ and $\omega_2$  depend on the adopted $L_{{\rm min}}$-value is shown in Fig.S6(c). One can see from this figure that the systematic drift of the optimal ($\omega_1$, $\omega_2$) observed for smaller $L_{{\rm min}}$-values tends to stop around $L_{{\rm min}}\simeq 60-72$. Similar fits have also been made for other quantities, where similar quality of the fitting results are found.

 Although the fit with two real correction-to-scaling exponents yields satisfactory fit as shown in Fig.S6(b) for the correlation-length ratios and in Fig.S7 for the specific heat and the energy Binder ratio, this type of fit has a problem. Namely, the coefficients of the correction terms $a_1$ and $a_2$ in eq.(\ref{tworeal-fit}) tend to be quite large and opposite in sign. In fact, in case of the correlation-length ratio shown in Fig.S6(b), $a_1=-23.9$ and $a_2=24.7$, leading to the correction terms comparable to or even greater than the leading term of unity, and a subtle cancellation between these two large correction terms takes account of the significant scaling correction. In fact, the same situation arises not only for the spin correlation-length ratio but also for other quantities. For example, the fit of the specific heat yields even greater correction-term coefficients of opposite sign, {\it i.e.\/}, $a_1=-78$ and $a_2=95$. We feel that such a correction is pathological, or at least not natural, and are lead to examine the second possible form of the correction described by a complex $\omega$.

\subsection{3. Analysis with a complex correction exponent} 

 In this subsection, we examine the correction with a complex correction-to-scaling exponent $\omega=\omega_R+i\omega_I$, which corresponds to the focus-type RG fixed point. In this case, the finite-size scaling form at $T=T_c$ is expected to take the form,
\begin{equation}
X \approx  L^x(1+a L^{-\omega_R} \cos(\omega_I \ln L + \phi)),
\label{complex-fit}
\end{equation}
where $\phi$ is a phase factor. Again, we begin our analysis with the correlation-length ratios without the exponent $x$ in its scaling form of eq.(\ref{complex-fit}). As was the case in the previous subsection, the fit based on eq.(\ref{complex-fit}) leads to many local minima with comparable $\chi^2$-values, and care has to be taken not to miss the true minima with the optimal $\chi^2$-value. Again, we perform the combined fit for $\xi_s^{\parallel}/L$, $\xi_s^{\perp}/L$, $\xi_\kappa^{\parallel}/L$ and $\xi_\kappa^{\perp}/L$ for various values of $L_{{\rm min}}$. In Fig.S8, we show all the local minima obtained by fitting the data of the correlation-length ratios $\xi_s^{\parallel,\perp}/L$ and $\xi_\kappa^{\parallel,\perp}/L$ in the $\omega_R$ vs. $\omega_I$ plane for $L_{{\rm min}}=60$, in which the color of data points represents the associated $\chi^2$-value. As can be seen from Fig.S8, there indeed exist many local minima in the fit. The best fit is obtained at $\omega_R\simeq 0.06$ and $\omega_I\simeq 0.63$, and the resulting fitting curves of $\xi_s^{\parallel,\perp}/L$ and $\xi_\kappa^{\parallel,\perp}/L$ are shown in Fig.11(a). In this optimal plot, the coefficient of the correction term $a$ of eq.(\ref{complex-fit}) has turned out to be $\simeq 0.3$, being free from the pathology we encountered in the case of the two real correction exponents.

\begin{figure}
  \begin{center}
    \includegraphics[width=\hsize]{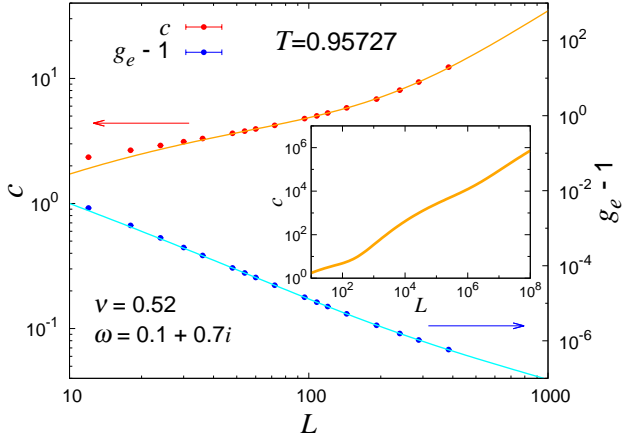}
   \caption{
(Color online) The scaling plots of the specific heat and the energy Binder ratio at the transition temperature $T=T_c=0.95727$. The scaling employed includes the correction term with a complex correction-to-scaling exponent, eq.(\ref{complex-fit}), with $L_{{\rm min}}=60$. The inset represents the expected behavior of the fitted functional form of $c$ at $T_c$ beyond the simulated lattice size.
}   
\end{center}
\end{figure}

 If $L_{{\rm min}}$ is varied, the resulting best values of $\omega_R$ and $\omega_I$ vary somewhat. How these best values of $\omega_R$ and $\omega_I$ depend on the adopted $L_{{\rm min}}$-value is shown in Fig.11(b). One can see from this figure that the systematic drift of the optimal ($\omega_R$, $\omega_I$) observed for smaller $L_{{\rm min}}$-values tends to stop around $L_{{\rm min}}\simeq 60-72$: See the dashed circle in the figure. Further increase of $L_{{\rm min}}$ beyond $L_{{\rm min}}=72$ means less number of available data points in the fit, leading to larger error bars. Hence, we judge that the choice of $L_{{\rm min}}=60$ or 72 would be optimal for the exponent estimate. The $\chi^2$/DOF of the fit turns out to be smallest for $L_{{\rm min}}=60$ with $\chi^2$/DOF=1.33, but not much different from that for $L_{{\rm min}}=72$, $\chi^2$/DOF=1.44. If we choose $L_{{\rm min}}=72$, we get $\omega_R\simeq 0.10$ and $\omega_I\simeq 0.68$, rather similar values to the $\omega$-value obtained for $L_{{\rm min}}=60$. Based on these observations, we set $\omega_R=0.1$ and $\omega_I=0.7$ in our following analysis.
\begin{figure}
  \begin{center}
    \includegraphics[width=\hsize]{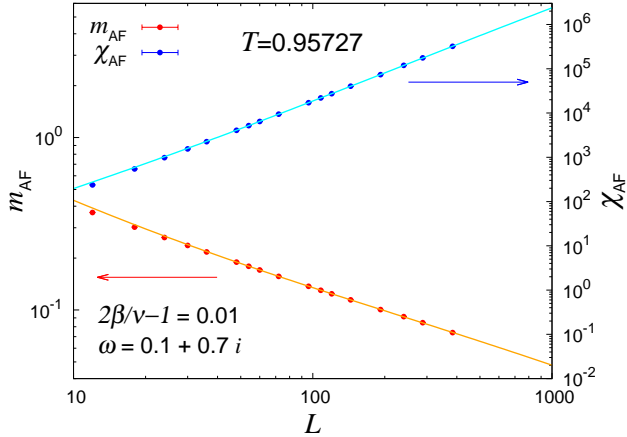}
   \caption{
(Color online) The scaling plots of the AF order parameter $m_{AF}$ and the AF susceptibility $\chi_{AF}$ at the transition temperature $T=T_c=0.95727$. The scaling employed includes the correction term with a complex-valued correction-to-scaling exponent, eq.(\ref{complex-fit}), with $L_{{\rm min}}=60$. 
} 
  \end{center}
\end{figure}
\begin{figure}
  \begin{center}
    \includegraphics[width=\hsize]{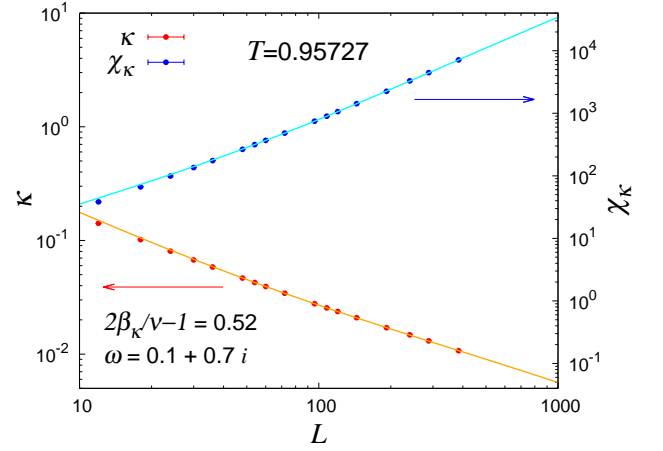}
   \caption{(Color online) The scaling plots of the chirality $\kappa$ and the chiral susceptibility $\chi_\kappa$ at the transition temperature $T=T_c=0.95727$. The scaling employed includes the correction term with a complex-valued correction-to-scaling exponent, eq.(\ref{complex-fit}), with $L_{{\rm min}}=60$. 
} 
  \end{center}
\end{figure}
 Similar finite-size-scaling fit has also been made for other quantities  at $T=T_c$ by using the $\omega_R$ and $\omega_I$ values determined above, {\it i.e.\/}, $\omega_R=0.1$ and $\omega_I=0.7$.

 The exponent $\nu$ is determined from the combined fit of the specific heat $c$ and the energy Binder ratio $g_e$. For $L_{{\rm min}}=60$, we get $\nu=0.52\pm 0.01$, and the resulting scaling plots are given in Fig.12. The coefficient of the correction term $a$ of the specific heat comes around $\simeq 0.9$, which seems to be a reasonable value. If we choose $L_{{\rm min}}=72$, we get $\nu=0.51\pm 0.01$, Overall, the fit turns out to reasonably reproduce the non-trivial size dependence of the specific heat and the energy Binder ratio.

 Although our largest size $L=384$ is already quite large, the issue of how the asymptotic size dependence described by eq.(\ref{complex-fit}) looks like in the still larger $L$-region might be interesting. Thus, in the inset of Fig.12, we show the asymptotic size dependence of the specific heat expected from the best fit of our MC data to eq.(\ref{complex-fit}) up to the size $L\simeq 10^8$. Oscillatory behavior is visible there, though the size required to clearly see such an oscillation is unrealistically large. 

 Putting reliable error bars on the estimates of $\omega_R$ and $\omega_I$ is rather difficult, since many local minima shown in Fig.S8 give comparable $\chi^2$-values. Here, we estimate the error bars of $\omega_R$ and $\omega_I$ based on the criterion of either (i) the local minimum no longer appearing in the fit, or (ii) the coefficient of the correction term $a$ for the specific heat exceeding five. Then, we get $\omega_R=0.1^{+0.4}_{-0.05}$ and $\omega_I=0.7^{+0.1}_{-0.4}$.

 The scaling plots of $m_{{\rm AF}}$ and $\chi_{{\rm AF}}$ are shown in Fig.13 for $L_{{\rm min}}=60$, where the best value of the exponent $\eta$ is determined  from the combined fit of these two quantities to be $\eta=0.01\pm 0.03$. If we choose  $L_{{\rm min}}=72$, we get $\eta=0.03\pm 0.06$.

 Similarly, the scaling plots of $\kappa$ and $\chi_{\kappa}$ are shown in Fig.14 for $L_{{\rm min}}=60$, where the best value of the exponent $\beta_\kappa/\nu$ is determined to be $2\beta_\kappa/\nu-1=0.52\pm 0.04$ from the combined fit of these two quantities. If we choose  $L_{{\rm min}}=72$, we get $2\beta_\kappa/\nu-1=0.52\pm 0.08$.

 In Fig.15, the $L_{{\rm min}}$-dependence of the exponents $\nu$, $2\beta/\nu-1=\eta$ and $2\beta_\kappa/\nu-1$ are shown. As mentioned, on the basis of our observation on the correction-to-scaling exponent $\omega$, we regard $L_{{\rm min}}=60$ or 72 as optimal. Then, $\nu$ is slightly greater than 0.5, and $\eta$ is slightly positive. As our final estimate of the exponents, we take a mean of the estimates for $L_{{\rm min}}=60$ and 72, and we get
\begin{eqnarray}
\nu=0.52(1),\ \ \eta&=&0.02(5),\ \ \beta_\kappa/\nu=0.76(6), \\
\omega&=&0.1^{+0.4}_{-0.05} + i\ 0.7^{+0.1}_{-0.4}.
\end{eqnarray}
With use of the scaling and hyperscaling relations, we get
\begin{eqnarray}
\alpha=0.44(3),\ \ \beta=0.26(2),\ \ \gamma=1.03(5), \\
\beta_\kappa=0.40(3),\ \ \gamma_\kappa=0.77(6).
\end{eqnarray}
\begin{figure}
  \begin{center}
    \includegraphics[width=\hsize]{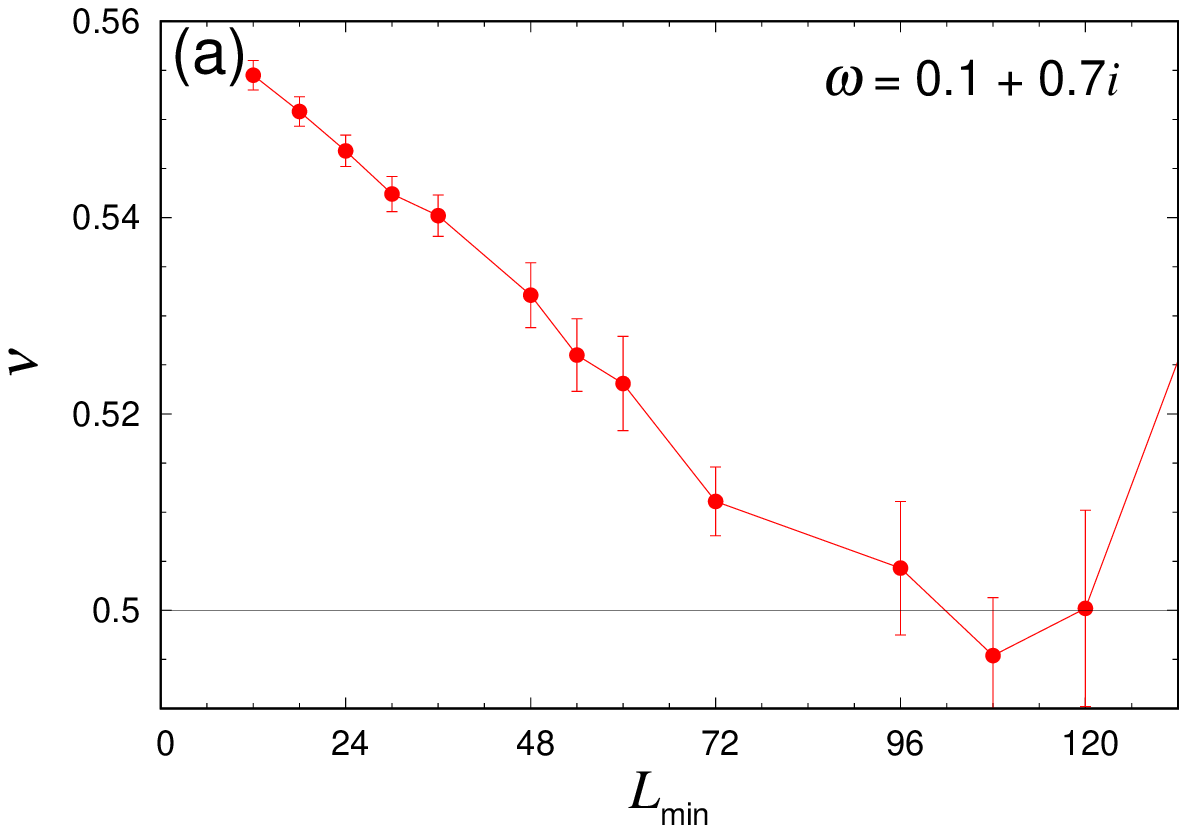}
    \includegraphics[width=\hsize]{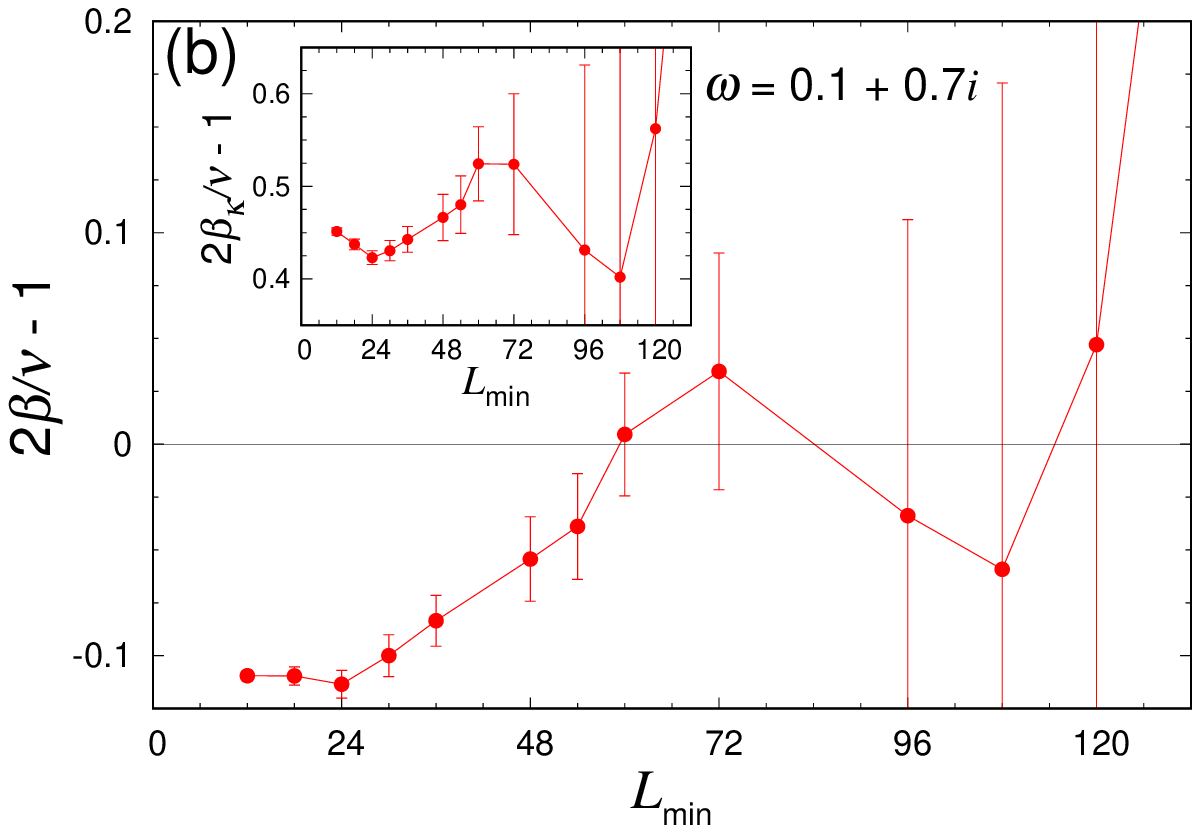}
   \caption{
(Color online) The exponents (a) $\nu$, and (b) $2\beta/\nu-1=\eta$, obtained by the scaling fit with the correction term with a complex-valued correction-to-scaling exponent, are plotted versus $L_{{\rm min}}$, the minimum lattice size used in the fit. The inset of (b) is the corresponding plot for the chirality exponent $2\beta_\kappa/\nu-1$.
} 
  \end{center}
\end{figure}

One can see that the obtained exponents $\nu$ and $\eta$ are close to the mean-field tricritical values of $\nu=0.5$ and $\eta=0$ governed by the Gaussian FP. By contrast, the chirality exponent $\beta_\kappa/\nu=0.76(6)$ is far from the mean-field tricritical value of unity, indicating that the criticality of the present model is not of mean-field tricritical governed by the Gaussian FP. Thus, the criticality as realized in the present model is not a trivial one, but is a highly nontrivial one, a chiral FP which is likely to be the focus-type FP. 

 With use of the estimate $\nu\simeq 0.52$, the exponent describing the size-dependence of the energy Binder ratio $g_e$, $\theta=2(3-\frac{1}{\nu})$, is estimated to be $\simeq 2.15$. In the inset of Fig.8(a), we estimate the effective $\theta$ from our $g_e$ data as a function of $L_{{\rm min}}$: It turned out to decrease from about 2.65 to 2.4 as $L_{{\rm min}}$ is increased to 192. Then, our present estimate of $\theta\simeq 2.15$ seems consistent with the MC data shown in Fig.8.

 As a consistency check, we also try to fit the $T_{{\rm peak}}(L)$-data shown in Fig.5 by the scaling form,
\begin{equation}
T_{{\rm peak}}=T_c + c'L^{1/\nu}(1+a'L^{-\omega_R}\cos(\omega_I \ln L+\phi)),
\label{Tpeak-fit}
\end{equation}
with $T_c=0,95727$, $\nu=0.52$, $\omega_1=0.1$ and $\omega_I=0.7$ where the non-universal constants $c'$, $a'$ and  $\phi$ are tuned for each $T_{{\rm peak}}(L)$. The resulting best fit has been given in Fig.5.

\section{VIII. Summary and discussion}

 We investigated the critical properties of the AF Heisenberg model on the 3D stacked-triangular lattice by means of a large-scale MC simulation in order to get insight into the controversial issue of the criticality of the noncollinear magnets with the $O(3)\times O(2)$ symmetry. The maximum size studied is $384^3$ considerably larger than the sizes studied by the previous numerical works on the model. Availability of such large-size data enabled us to examine the detailed critical properties, including the effect of corrections to the leading scaling. The transition temperature was located rather precisely as $T_c=0.957270\pm 0.000004$. We have obtained a strong numerical evidence of the continuous nature of the transition. The energy distribution always exhibits a single peak characteristic of a continuous transition at any temperature and for all sizes studied up to our largest size of $L=384$, in contrast to the previous report of the double-peak structure for $L=150$. In addition, on increasing $L$, the energy Binder ratio exhibits a behavior further deviating from the one expected for a first-order transition. Confirming the continuous nature of the transition, its critical properties are examined carefully on the basis of our extensive set of data. The existence of significant corrections to the leading scaling was indicated, and we performed a careful analysis by taking account of the possible corrections. We then get the estimates of critical exponents $\alpha=0.44(3)$, $\beta=0.26(2)$, $\gamma=1.03(5)$, $\nu=0.52(1)$, $\eta=0.02(5)$, and the chirality exponents $\beta_\kappa=0.40(3)$ and $\gamma_\kappa=0.77(6)$. 

We also obtained an indication that the underlying FP was of the focus-type, {\it i.e.\/}, we obtained the complex-valued  correction-to-scaling exponent, $\omega=0.1^{+0.4}_{-0.05} + i\ 0.7^{+0.1}_{-0.4}$. The focus-like nature of the chiral FP accompanied by the spiral-like RG flow is likely to be the origin of the apparently complicated critical behavior of the model we observed. Thus, we find numerical evidence of the existence of the $O(3)$ chiral (or $O(3)\times O(2)$) universality class governed by the new chiral FP.

 The focus-like FP and the associated ``oscillatory'' critical behavior might provide further interesting possibility. As shown in in Figs.1 and 2 of Ref.\cite{Calabrese1}, the RG flow around the focus-like chiral FP could move, upon renormalization, from the parameter region of a continuous transition into that of a first-order transition, and then get back to the continuous-transition region, eventually flowing into the chiral FP. If one looks at the energy distribution on various length scales under such circumstances, it would exhibit a single peak characteristic of a continuous transition for smaller system sizes, then exhibit double peaks characteristic of a first-order transition for larger sizes, but eventually exhibit a single peak characteristic of a continuous transition again for still larger sizes. Although we did not observe any double-peak structure for any lattice size in our present simulation on the stacked-triangular AF Heisenberg model, it might be interesting to point out that, for the stacked-triangular AF {\it XY\/} model, several MC simulations reported that a single-peak structure of the energy distribution observed for smaller lattices changed into the double-peak one for larger lattices, arguing that the transition should eventually be first-order \cite{Itakura,Peles,Kanki}. In view of the possible focus-like feature of the chiral FP, however, the possibility of the observed double-peak structure finally changing into the single-peak one should also be kept in mind.
 
 The exponents we obtained, especially $\nu$ (also related $\alpha$, $\beta$ and $\gamma$), differ somewhat from the corresponding values reported by the earlier MC simulations on the same model for smaller sizes ($L\leq 60$), though the continuous nature of the transition is common \cite{KawamuraMCH,KawamuraMCHXY,Bhattacharya,Mailhot,Loison}. This deviation is likely to be due to the large correction-to-scaling as described above, since the exponents of these earlier reports came close to the effective exponents we obtained for smaller lattices. For example, the effective $\nu$ we obtained for $L_{{\rm min}}\leq 60$ came around $0.58\lesssim \nu_{{\rm eff}}\lesssim 0.60$ as shown in Fig.10(a), while the estimate of Ref.\cite{KawamuraMCHXY} for $L\leq 60$ gave $\nu=0.59(2)$.

 Our present estimates of exponents come rather close to those of the six-loop perturbative massive RG calculation $\alpha=0.35(9)$, $\beta=0.30(2)$, $\gamma=1.06(5)$, $\nu=0.55(3)$, $\eta=0.073(94)$, $\beta_\kappa=0.38(10)$ and $\gamma_\kappa=0.89(10)$, while differ somewhat from those of the five-loop massless RG calculation $\alpha=0.11(15)$, $\beta=0.34(3)$, $\gamma=1.20(8)$, $\nu=0.63(5)$, $\eta=0.08(3)$, $\beta_\kappa=0.54(17)$ and $\gamma_\kappa=0.81(23)$, and those of the conformal-bootstrap calculation $\alpha=0.10(5)$, $\beta=0.34(1)$, $\gamma=1.22(3)$, $\nu=0.63(2)$, $\eta=0.078(6)$, $\beta_\kappa=0.56(2)$ and $\gamma_\kappa=0.77(3)$, though the continuous nature of the transition is also in common.

 In fact, our estimates of $\nu=0.52(1)$ and $\eta=0.02(5)$ are quite close to the mean-field tricritical value governed by the Gaussian FP. Since the chirality exponents $\beta_\kappa=0.40(3)$ and $\gamma_\kappa=0.77(6)$ largely differ from the corresponding mean-field tricritical values $\beta_\kappa=1/2$ and $\gamma_\kappa=1/2$, which can be derived from the chiral-crossover exponent $\phi_\kappa=1$ and $\alpha=1/2$ at the Gaussian FP, the chiral FP cannot be the standard Gaussian FP. Furthermore, the Gaussian FP is strongly unstable with respect to the two quartic couplings of the $O(3)\times O(2)$ LGW Hamiltonian, and practically is inaccessible. At present, we do not know whether the closeness of the obtained exponents $\nu=0.52(1)$ and $\eta=0.02(5)$ to the mean-field tricitical values is just accidental, or has a deeper reason behind that. Numerically, it is for sure that the chiral FP is not the standard Gaussian FP.

 The possible focus-like feature of  the chiral FP is consistent with the suggestion from the higher-order perturbative RG including both the MZM \cite{Calabrese1} and the $\overline{{\rm MS}}$ \cite{Calabrese2} schemes. By contrast, the conformal-bootstrap analysis assumes the absence of the focus point \cite{Nakayama}, and there still remains a problem. Our estimate of the complex-valued correction-to-scaling exponent $\omega=0.1^{+0.4}_{-0.05} + i\ 0.7^{+0.1}_{-0.4}$  is to be compared with the corresponding estimates from the perturbative RG calculations, {\it i.e.\/},  $\omega=1.00(20) + i\ 0.80(25)$ from the six-loop MZM, and  $\omega=0.9(4) + i\ 0.7(3)$ from the five-loop $\overline{{\rm MS}}$. Though the imaginary part agrees well with each other, our estimate of the real part came smaller than the RG estimates.

 Our present result indicating a continuous transition is in contrast to the functional RG result, which invariably suggests a first-order transition \cite{Tisser1,Tisser2,Delamotte1,Delamotte2,Delamotte3}. The issue of why the nonperturbative functional RG and the perturbative RG at $d=3$ yield different answers has remained controversial and needs to be understood. The present MC result basically support the perturbative $d=3$ RG and the conformal bootstrap results, the perturbative RG result based on the massive MZM scheme, in particular.

 Our MC result sharply contradicts the report of a first-order transition for the same model by Ref.\cite{Diep} on the basis of the Wang-Landau method. The energy distribution computed in our present calculation always exhibited a single peak for any size in the range $12\leq L\leq 384$, in sharp contrast to the double peaks observed in Ref.\cite{Diep} for the sizes $L=120$ and 150. Hence, the possible spiral-like RG flow discussed above cannot be invoked as a resolution of the observed discrepancy.  Our data other than the energy distribution do not exhibit any sign of a first-order transition up to the size $L=384$. We do not know the reason why Ref.\cite{Diep} observed a double-peak structure in their data of the energy distribution, but just suspect there might be something wrong in the application of the Wang-Landau method.

 While we did not go into details about the experimental connection in the present paper, experimental situations in the last century were extensively reviewed in Ref.\cite{Kawamura-review}, and we believe that most of its contents remain effective even now. Overall, most of the experiments performed on the stacked-triangular antiferromagnets reported a continuous transition characterized by non-standard exponents distinct from the standard $O(n)$ values. 

 Sometimes, a weak first-order transition was claimed based on the experimental observation of the deviation from the ideal power-law scaling relation in the temperature range close to $T_c$, not on the direct observation of the nonzero latent heat nor on the clear discontinuity in physical quantities \cite{Kawamura-review}. The deviation from the ideal power-law, however, could arise from various sources. The oscillatory critical behavior due to the complex-valued correction-to-scaling exponent might occur as perturbative RG computations at $d=3$ and our present calculation suggested. Furthermore, in real materials, non-ideal sources might also come into play causing the deviation from the ideal critical behavior, {\it e.g.\/}, the inevitably existing randomness like defects and impurities, the temperature inhomogeneity in the sample, {\it etc\/}. Hence, in order to experimentally establish the first-order nature of the transition, one should probe a sharp discontinuity such as the nonzero latent heat. In addition, even if the first-order transition would have been established in a few materials, it does not automatically guarantee that the chiral FP does not exist in nature, simply because, even in the presence of the stable FP, certain systems can still exhibit a first-order transition depending on the microscopic details of the system, when the bare parameters describing that system lie outside the domain of attraction of the chiral FP. 

 Although our present analysis has given strong numerical evidence of the continuous nature of the noncollinear transition of frustrated Heisenberg magnets, it still does no completely rule out the possibility of an extremely weak first-order transition in the mathematical sense. However, such a hypothetical first-order transition should be extremely weak, visible only on the length scale considerably longer than our present largest size of $L=384$, which is already quite long. If one translates the length scale into the (reduced) temperature scale assuming the correlation-length relation $\xi = |(T-T_c)/T_c|^{-\nu}$ with our present estimate $\nu\simeq 0.52$, $\xi\sim L= 384$ means $|(T-T_c)/T_c|\simeq 10^{-5}$, quite a small number usually uncontrollable in experiments. Of course, the system size available in real experiments could be longer than $L=384$, but in reality such macroscopic samples suffer from the randomness or inhomogeneity such as defects and impurities which would modify or round the transition behavior at close vicinity of $T_c$. Indeed, defects or impurities at every 384 sites already means their density of order $10^{-8}$. In this sense, we might already be reaching the limit of the experimentally accessible critical regime. Even if the transition might eventually become very weakly first-order beyond this length scale, it may largely be a purely academic matter.

 The physically important thing is that, as the anomalous crical behavior has certainly been onbserved both experimentally and numerically in a variety of frustrated noncollinear magnets on the already quite long length-scale $L\lesssim 384$, the nature and the origin of it should be explained and understood. Setting aside a largely academic issue of whether the transition being either continuous or extremely weakly first-order beyond the length scale $L\gtrsim 384$, we definitely need the physical understanding of the anomalous critical behavior observed in many experiments and model simulations. The picture emerging from our present calculation is that the transition is continuous characterized by the focus-like chiral FP. While this picture seems well consistent with our present MC data and with experiments, we do not know for sure whether it is the only and most effective description of the anomalous critical behavior observed experimentally and numerically on the length scale of $L\lesssim 384$. 

 Some of open questions might be: (i) The noncollinear criticality really exhibits the focus-like critical behavior ? (ii) If it does, how it reconciles with the conformal-bootstrap theory ? (iii) Why the thermal and magnetic exponents are close to the mean-field-tricritical values in spite of the chiral FP being not the standard Gaussian FP ? (iv) Why various RG schemes which give mutually consistent answers in the standard $O(n)$ problem give mutually different and sometimes even contradicting answers in the $O(n)\times O(2)$ problem ? {\it etc. etc\/}.  These issues might still remain to be challenging, and the issue of the noncollinear or $O(n)\times O(m)$ criticality most probably contains rich physics in it, providing an important key to make progress in the challenge.

\begin{acknowledgements}
The authors are thankful to Dr. Y. Nakayama and Dr. T. Ohtsuki for useful discussion and comments. This study was supported by JSPS KAKENHI Grants No. 17H06137. We are thankful to ISSP, the University of Tokyo, and to YITP, Kyoto University, for providing us with CPU time.
\end{acknowledgements}

\section{Appendix. Derivation of the exponent describing the size dependence of the energy Binder ratio $g_e$}

In this appendix, we give the derivation of eq.(\ref{ge}) of the main text describing the size dependence of the energy Binder ratio $g_e$. In terms of the energy per spin $e$, we introduce the quantities $\delta_n$ ($n=2,3,4,\cdots$) by
\begin{eqnarray}
\delta_n=\frac{\langle (e-\langle e\rangle)^n\rangle}{\langle e\rangle^n}.
\end{eqnarray}
The energy Binder ratio $g_e$ can be rewitten in terms of $\delta_n$ as
\begin{eqnarray}
g_e=\frac{1+6\delta_2+4\delta_3+\delta_4}{(1+\delta_2)^2}.
\end{eqnarray}
Since $1\gg \delta_2 \gg \delta_3 \gg \delta_4$, one has
\begin{eqnarray}
g_e-1\simeq 4\delta_2.
\end{eqnarray}
Now, $\delta_2$ can be written in terms of the internal energy per spin $\bar e$ and the specific heat per spin $c$ as
\begin{eqnarray}
\delta_2=\frac{c}{N \bar e^2}.
\end{eqnarray}
At $T=T_c$, the leading size dependence of $\bar e$ and $c$ are expected to be
\begin{eqnarray}
\bar e = \bar e_0 + {\rm const.}\times L^{\frac{\alpha-1}{\nu}} + \cdots, \\
c = c_0 + {\rm const.}\times L^{\frac{\alpha}{\nu}} + \cdots, 
\end{eqnarray}
where $\bar e_0$ and $c_0$ are non-singular constants. Note that, since $\alpha<1$ at the continuous transition, the regular term $\bar e_0$ gives the leading contribution to the $L$-dependence of the energy $\bar e$. In case of the specific heat $c$, the leading contribution to the $L$-dependence comes from the singular second term $L^{\frac{\alpha}{\nu}}$ if $\alpha>0$ as in the present case, while it comes from the non-singular constant term $c_0$ if $\alpha<0$. Then, the leading contribution to the $L$-dependence of $\delta_2$ and $g_e-1$ should be given by
\begin{eqnarray}
g_e-1&\approx& L^{-2(d-\frac{1}{\nu})} = L^{-2(3-\frac{1}{\nu})},\ \  \alpha>0, 
\label{ge2}
\\  
g_e-1&\approx& L^{-d} = L^{-3},\ \  \alpha<0.
\end{eqnarray}
In the present case, eq.(\ref{ge2}) should hold since $\alpha>0$.

\end{document}